\documentclass[prd,twocolumn,floatfix,amsmath,nofootinbib,amssymb,superscriptaddress]{revtex4}
\usepackage{graphicx,color,dcolumn,booktabs,bm}
\usepackage{longtable,lscape}
\usepackage{pdfpages}
\usepackage{txfonts}
\usepackage{overpic}
\usepackage{amssymb}
\usepackage{makecell}
\usepackage{indentfirst}
\usepackage{feynmf}
\usepackage{slashed}
\usepackage[utf8]{inputenc}
\usepackage{cases}
\usepackage{color}
\usepackage{multirow}
\usepackage{threeparttable}
\usepackage{epstopdf}
\usepackage{enumerate}
\usepackage{subfigure}
\usepackage{diagbox}
\usepackage{graphicx,color,dcolumn,booktabs,bm}
\usepackage{mathrsfs}
\usepackage{cancel}
\usepackage{float}
\usepackage[misc]{ifsym}
\usepackage[colorlinks,
citecolor=blue,
anchorcolor=red,
menucolor=red,
linkcolor=red,
filecolor=red,
runcolor=red,
urlcolor=blue,
frenchlinks=red]{hyperref}
\usepackage{array}
\usepackage{booktabs}
\usepackage{amsmath}
\usepackage{slashed}
\begin{document}
\title{Analysis of the two-body strong decays of the hidden-charm pentaquark states in QCD sum rules}
\author{Zi-Hao Chen}
\affiliation{Key Laboratory of Nuclear Data, China Institute of Atomic Energy, Beijing, 102413, P. R. China}
\affiliation{Department of Physics, North China Electric Power University, Baoding 071003, P. R. China}

\author{Han-Xiong Huang}
\email{hhanxiong@163.com}
\affiliation{Key Laboratory of Nuclear Data, China Institute of Atomic Energy, Beijing, 102413, P. R. China}

\author{Jie Lu}
\email{l17693567997@163.com}
\affiliation{School of Physics, Southeast University, Nanjing 210094, P. R. China}

\author{Xiao-Song Yang}
\affiliation{Frontiers Science Center for Rare Isotopes, and School of Nuclear Science and Technology, Lanzhou University, Lanzhou 730000, P. R. China}

\author{Zhi-Gang Wang}
\email{zgwang@aliyun.com}
\affiliation{Department of Physics, North China Electric Power University, Baoding 071003, P. R. China}

\author{Guo-Liang Yu}
\email{yuguoliang2011@163.com}
\affiliation{Department of Physics, North China Electric Power University, Baoding 071003, P. R. China}
\date{\today }
\begin{abstract}
In the present work, we study the two-body strong decays of the hidden-charm pentaquark states with the quark content $uudc\bar c$ and the quantum numbers $I(J^P)=\frac{1}{2}(\frac{1}{2}^-)$ in the framework of the three-point QCD sum rules. The initial pentaquark states are described by four local diquark-diquark-antiquark type interpolating currents with definite isospin. We construct the three-point correlation functions for the decay channels $P_c\to \eta_c p$, $J/\psi p$, $\Lambda_c\bar D$, $\Lambda_c\bar D^{*}$ and $\Sigma_c\bar D$, and derive the corresponding QCD sum rules for the strong coupling constants. At the hadron side, the correlation functions are expressed in terms of the hadron masses, pole residues, decay constants and strong coupling constants. At the QCD side, they are calculated by carrying out the operator product expansion with the full quark propagators, where the vacuum condensates up to dimension 10 are taken into account. After matching the two representations and performing the double Borel transformations, we extract the strong coupling constants from the selected Lorentz structures. With the obtained coupling constants, we evaluate the partial decay widths and discuss the possible assignments of the corresponding pentaquark states. The numerical results indicate that two of the compact hidden-charm pentaquark states can be related to the $P_c(4312)$ and $P_c(4457)$, respectively, while the other two lower-mass states may be regarded as possible hidden-charm pentaquark candidates to be searched for in future experiments. The present results may be useful for identifying the hidden-charm pentaquark states in future experiments.
\end{abstract}

\maketitle

\section{Introduction}\label{sec1}

In 2015, the LHCb Collaboration observed two hidden-charm pentaquark candidates $P_c(4380)$ and $P_c(4450)$, in the $J/\psi p$ invariant mass spectrum of the $\Lambda_b^0\to J/\psi K^-p$ decay with a statistical significance larger than $9\sigma$~\cite{LHCb:2015yax}. Their Breit-Wigner masses and widths are reported as:
\begin{eqnarray}
\notag
\Big(m,\Gamma\Big)_{P_c(4380)}&&=\Big(4380\pm8\pm29,205\pm18\pm86\Big)~\mathrm{MeV}, \\
\notag
\Big(m,\Gamma\Big)_{P_c(4450)}&&=\Big(4449.8\pm1.7\pm2.5,39\pm5\pm19\Big)~\mathrm{MeV}.
\end{eqnarray}
The preferred spin-parity assignments for these two states were $J^P=\frac{3}{2}^{-}$ and $\frac{5}{2}^{+}$, respectively, while the opposite-parity assignments could not be excluded. 

In 2019, using a data sample about one order of magnitude larger than the previous one, the LHCb Collaboration observed a narrow structure $P_c(4312)$ and resolved the $P_c(4450)$ peak into another two narrow structures $P_c(4440)$ and $P_c(4457)$~\cite{LHCb:2019kea}. The measured masses and widths are
\begin{eqnarray}
\notag
\Big(m,\Gamma\Big)_{P_c(4312)}&&=\Big(4311.9\pm0.7^{+6.8}_{-0.6},9.8\pm2.7^{+3.7}_{-4.5}\Big)~\mathrm{MeV},\\
\notag
\Big(m,\Gamma\Big)_{P_c(4440)}&&=\Big(4440.3\pm1.3^{+4.1}_{-4.7},20.6\pm4.9^{+8.7}_{-10.1}\Big)~\mathrm{MeV},\\
\notag
\Big(m,\Gamma\Big)_{P_c(4457)}&&=\Big(4457.3\pm0.6^{+4.1}_{-1.7},6.4\pm2.0^{+5.7}_{-1.9}\Big)~\mathrm{MeV}.
\end{eqnarray}
Up to now, the spin and parity quantum numbers of the three narrow structures have not been determined experimentally. 

In 2021, the LHCb Collaboration reported a new structure $P_c(4337)$ in the $J/\psi p$ and $J/\psi \bar p$ invariant mass spectra of the $B_s^0\to J/\psi p\bar p$ decay~\cite{LHCb:2021chn}. The statistical significance depends on the assumed spin-parity assignment and lies in the range from $3.1-3.7\sigma$. Its mass and width are
\begin{eqnarray}
\notag
\Big(m,\Gamma\Big)_{P_c(4337)}&&=\Big(4337^{+7+2}_{-4-2},29^{+26+14}_{-12-14}\Big)~\mathrm{MeV}.
\end{eqnarray}
As these $P_c$ structures are observed in the $J/\psi p$ channel, their minimal valence-quark content is $uudc\bar c$. They are therefore natural hidden-charm pentaquark candidates. The experimental progress and theoretical interpretations of the hidden-charm pentaquark and related exotic states have been reviewed extensively in Refs.~\cite{Chen:2016qju,Liu:2019zoy,Brambilla:2019esw,Wang:2025sic}.

The observed $P_c(4312)$, $P_c(4380)$, $P_c(4440)$ and $P_c(4457)$ are close to the thresholds of $\Sigma_c\bar{D}$, $\Sigma_c^*\bar{D}$, $\Sigma_c\bar{D}^*$ and $\Sigma_c\bar{D}^*$, respectively. This feature has motivated many interpretations in terms of corresponding hadronic molecular states, and their mass spectra, production and decay properties have been investigated extensively in the molecular picture~\cite{Guo:2017jvc,PavonValderrama:2019nbk,Liu:2019zvb,Wang:2019nwt,Du:2019pij,He:2019ify,Wang:2015qlf,Lu:2016nnt,Azizi:2018bdv,Lin:2019qiv,Voloshin:2019aut,Gutsche:2019mkg,Ling:2021lmq,He:2019rva,Wu:2019rog,Xiao:2019mvs,Ke:2019bkf,Chen:2020pac,Dong:2020nwk,Xu:2020flp,Deng:2026gqe}. However, a mass close to a hadronic threshold does not necessarily imply a pure molecular configuration. Compact pentaquark configurations may also appear in the same energy region~\cite{Weng:2019ynv,Wang:2019dsi,Stancu:2021rro,Wang:2022fdu,Zhu:2019iwm,Yamaguchi:2019seo,Shi:2021wyt}. Moreover, threshold cusps, triangle singularities and coupled-channel effects may also play a role in shaping the observed structures~\cite{Du:2019pij,Guo:2019kdc,Eides:2019tgv,Sakai:2019qph,Kuang:2020bnk}. In addition, for the $P_c(4337)$ state, the mass is close to several relevant thresholds, but it cannot be assigned to a single baryon-meson threshold directly. Therefore, both the hadronic molecular scenario and the compact pentaquark scenario deserve further investigations~\cite{LHCb:2021chn,Guo:2019kdc,Eides:2019tgv,Kuang:2020bnk,Wang:2025qtm}.

As a powerful non-perturbation approach, the QCD sum rules is time-honored and widely used in studying the properties of hadrons~\cite{Shifman:1978bx,Shifman:1978by,Reinders:1984sr,Colangelo:2000dp}. In recent years, this approach have been successfully extended to the study of mass spectra and decay properties of multiquark states~\cite{Albuquerque:2018jkn,Wang:2020cme,Wang:2019hyc,Wang:2020rdh,Wang:2023kir,Chen:2026ybf}, and the QCD sum rule analyses of hidden-charm pentaquark currents have been carried out in both color singlet-color singlet molecular configurations and compact diquark-diquark-antiquark configurations~\cite{Wang:2015epa,Wang:2019got,Wang:2015ava,Wang:2015ixb,Chen:2016otp,Wang:2015wsa,Wang:2018waa,Wang:2025fqh,Wang:2025pjt,Wang:2026dqi,Wang:2026thx}. Our previous work systematically calculated the mass spectra of these $P_c$ states in the molecular configuration by strictly distinguishing isospin with the two-point QCD sum rules~\cite{Wang:2022ltr}. The $P_c(4312)$, $P_c(4380)$, $P_c(4440)$ and $P_c(4457)$ were tentatively assigned as $\Sigma_c\bar{D}$, $\Sigma_c^*\bar{D}$, $\Sigma_c\bar{D}^*$ and $\Sigma_c^*\bar{D}^*$ molecular states with $I(J^P)=\frac{1}{2}(\frac{1}{2}^{-})$, $\frac{1}{2}(\frac{3}{2}^{-})$, $\frac{1}{2}(\frac{3}{2}^{-})$ and $\frac{1}{2}(\frac{5}{2}^{-})$, respectively. Next, we used the three-point QCD sum rules to calculate the decay widths of these states~\cite{Wang:2023ews,Wang:2026bls}, the predicted total widths $10.89^{+8.97}_{-8.09}$, $158.86^{+12.67}_{-6.82}$, $17.37^{+3.68}_{-3.68}$, $5.63^{+0.85}_{-0.85}$ MeV for $P_c(4312)$, $P_c(4338)$, $P_c(4440)$ and $P_c(4457)$ are compatible with the experimental values within uncertainties, which indicates that the molecular assignment can provide a possible description of these $P_c$ states. Although these observed $P_c$ states can be reasonably explained within the molecular state configuration, this does not exclude the compact pentaquark configurations or other interpretations. We hope to systematically analyze the mass spectra and decay properties of these $P_c$ states under the compact pentaquark configuration in order to further understand the internal structure of these states.

Recently, the mass spectrum of compact diquark-diquark-antiquark type $uudc\bar c$ pentaquark states with definite isospin $I=\frac{1}{2}$ was studied in QCD sum rules by constructing interpolating currents with explicit isospin structures~\cite{Wang:2025qtm}. For the four interpolating currents $J_1-J_4$ with $I(J^P)=\frac{1}{2}(\frac{1}{2}^{-})$,
the predicted masses are
\begin{eqnarray}
\notag
M_{P_{c1}}&&=4.31\pm0.11~\mathrm{GeV}, \\
\notag
M_{P_{c2}}&&=4.45\pm0.11~\mathrm{GeV}, \\
\notag
M_{P_{c3}}&&=4.20\pm0.11~\mathrm{GeV}, \\
\notag
M_{P_{c4}}&&=4.25\pm0.11~\mathrm{GeV}.
\end{eqnarray}
The mass of the $P_{c1}$ state coupled to $J_1$ is consistent with the experimental mass of the $P_c(4312)$, while the mass of the $P_{c2}$ state coupled to $J_2$ is compatible with those of the $P_c(4440)$ and $P_c(4457)$. These results suggest that compact pentaquark configurations may provide an alternative description of the observed $P_c$ structures. The states coupled to $J_3$ and $J_4$ have lower central masses; in particular, the lowest state lies slightly above the $\bar D\Lambda_c$ threshold. Since the uncertainties of the predicted masses are sizeable and the mass regions of different states overlap with each other, the mass spectrum alone is not sufficient to make a reliable assignment.

It is therefore necessary to study the two-body strong decays of these states. In this article, we construct the three-point correlation functions for the above strong decay processes and investigate the strong decay properties of the hidden-charm pentaquark states with the quark content $uudc\bar c$ and the quantum numbers $I(J^P)=\frac{1}{2}(\frac{1}{2}^{-})$ in the framework of the three-point QCD sum rules. We consider the possible decay channels
\begin{eqnarray}
\notag
P_{ci} &\to \eta_c p,~~J/\psi p,~~\Lambda_c\bar D,~~\Lambda_c\bar D^{*},~~\Sigma_c\bar D,~~\mathrm{and}~~\Sigma_c\bar D^{*},
\end{eqnarray}
and retain only the channels allowed by the available phase space in the numerical analysis.

This article is arranged as follows. After introduction in Sec.~\ref{sec1}, we derive the QCD sum rules for the strong coupling constants in Sec.~\ref{sec2}. In Sec.~\ref{sec3}, we present the numerical results and discussions. Sec.~\ref{sec4} is reserved for the conclusion.

\section{QCD sum rules for the strong decays of the hidden-charm pentaquark states}\label{sec2}
Firstly, we specify the interpolating currents for the initial hidden-charm pentaquark states. In the present work, we focus on the compact diquark-diquark-antiquark configurations with the quark content $uudc\bar c$. The local pentaquark currents with the definite isospin $I=\frac{1}{2}$ have been constructed in Ref.~\cite{Wang:2025qtm}, where the isospin structures of the light quarks were treated explicitly. The interpolating currents for the four spin-parity $\frac{1}{2}^-$ states are as follows,
\begin{eqnarray}\label{eq:1}
\notag
J_1(x) &&=\varepsilon_{ila}\varepsilon_{ijk}\varepsilon_{lmn}u^{jT}(x) \mathcal{C}\gamma_5 d^k(x)u^{mT}(x) \mathcal{C}\gamma_5 c^n(x)\mathcal{C}\bar c^{aT}(x), \\
\notag
J_2(x) &&=\varepsilon_{ila}\varepsilon_{ijk}\varepsilon_{lmn}u^{jT}(x) \mathcal{C}\gamma_5 d^k(x)u^{mT}(x) \mathcal{C}\gamma_\mu c^n(x)\gamma_5\gamma^\mu \mathcal{C}\bar c^{aT}(x), \\
\notag
J_3(x) &&=\frac{\varepsilon_{ila}\varepsilon_{ijk}\varepsilon_{lmn}}{\sqrt{2}}
\Big[u^{jT}(x) \mathcal{C}\gamma_\mu u^k(x)d^{mT}(x) \mathcal{C}\gamma^\mu c^n(x) \\
\notag
&&-u^{jT}(x) \mathcal{C}\gamma_\mu d^k(x)u^{mT}(x) \mathcal{C}\gamma^\mu c^n(x)\Big]\mathcal{C}\bar c^{aT}(x), \\
\notag
J_4(x) &&=\frac{\varepsilon_{ila}\varepsilon_{ijk}\varepsilon_{lmn}}{\sqrt{2}}
\Big[u^{jT}(x) \mathcal{C}\gamma_\mu u^k(x)d^{mT}(x) \mathcal{C}\gamma_5 c^n(x) \\
&&-u^{jT}(x) \mathcal{C}\gamma_\mu d^k(x)u^{mT}(x) \mathcal{C}\gamma_5 c^n(x)\Big]
\gamma_5\gamma^\mu \mathcal{C}\bar c^{aT}(x).
\end{eqnarray}
Here $\varepsilon_{ijk}$ is the 3 dimension Levi-Civita tensor, $i$, $j$, $k$, $l$, $m$, $n$ and $a$ are color indices, $\mathcal{C}$ denotes the charge conjugation matrix, and the superscript $T$ represents the transpose in the Dirac space. The currents $J_1(x)$ and $J_2(x)$ correspond to the $[ud][uc]\bar c$ configurations, while $J_3(x)$ and $J_4(x)$ correspond to the normalized isospin $\frac{1}{2}$ combinations $[uu][dc]\bar c-[ud][uc]\bar c$. The factor $1/\sqrt{2}$ is introduced to normalize the corresponding isospin wave functions.

These currents were used in Ref.~\cite{Wang:2025qtm} to study the mass spectrum of the compact hidden-charm pentaquark states. In the present work, we use them as the interpolating currents of the initial states and further study their two-body strong decays. Taking into account the conservation of isospin, angular momentum and parity, as well as the available phase space for the central masses obtained in Ref.~\cite{Wang:2025qtm}, we consider the following decay channels,
\begin{eqnarray*}
J_1: &&
P_{c1}\to \eta_c p,~J/\psi p,~\bar D\Lambda_c,~\bar D^{*}\Lambda_c;\\
J_2: &&
P_{c2}\to \eta_c p,~J/\psi p,~\bar D\Lambda_c,~\bar D^{*}\Lambda_c,~\bar D\Sigma_c;\\
J_3: &&
P_{c3}\to \eta_c p,~J/\psi p,~\bar D\Lambda_c;\\
J_4: &&
P_{c4}\to \eta_c p,~J/\psi p,~\bar D\Lambda_c.
\end{eqnarray*}
where $p$ denotes the proton. To extract the strong coupling constants for the above decay processes, we introduce the following three-point correlation functions,
\begin{eqnarray}\label{eq:2}
\notag
\Pi^{i}_{\eta_c p}(p,p') &&= i^2 \int d^4x d^4ye^{ip\cdot x}e^{iq\cdot y}\\
\notag
&&\times \langle 0|\mathcal{T}\{J_{\eta_c}(x)J_p(y)\bar J_{i}(0)\}|0\rangle ,\\
\notag
\Pi^{i}_{J/\psi p,\mu}(p,p') &&= i^2 \int d^4x d^4ye^{ip\cdot x}e^{iq\cdot y}\\
\notag
&&\times \langle 0|\mathcal{T}\{J_{J/\psi,\mu}(x)J_p(y)\bar J_{i}(0)\}|0\rangle ,\\
\notag
\Pi^{i}_{\Lambda_c\bar D}(p,p') &&= i^2 \int d^4x d^4ye^{ip\cdot x}e^{iq\cdot y}\\
\notag
&&\times \langle 0|\mathcal{T}\{J_{\Lambda_c}(x)J_{\bar D}(y)\bar J_{i}(0)\}|0\rangle, \\
\notag
\Pi^{i}_{\Lambda_c\bar D^*,\mu}(p,p') &&= i^2 \int d^4x d^4ye^{ip\cdot x}e^{iq\cdot y} \\
\notag
&&\times \langle 0|\mathcal{T}\{J_{\Lambda_c}(x)J_{\bar D^*,\mu}(y)\bar J_{i}(0)\}|0\rangle, \\
\notag
\Pi^{i}_{\Sigma_c\bar D}(p,p') &&= i^2 \int d^4x d^4y e^{ip\cdot x}e^{iq\cdot y}\\
&&\times \langle 0|\mathcal{T}\{J_{\Sigma_c}(x)J_{\bar D}(y)\bar J_{i}(0)\}|0\rangle .
\end{eqnarray}
Here, the subscribes $i=1\sim4$ are for different interpolating current and $\mathcal{T}$ denotes time-order production. The interpolating currents for the final state hadrons are chosen as
\begin{eqnarray}\label{eq:3}
\notag
J_{\eta_c}(x)&&= \bar c^{i'}(x)i\gamma_5 c^{i'}(x), \\
\notag
J_{J/\psi,\mu}(x)&&= \bar c^{i'}(x)\gamma_\mu c^{i'}(x), \\
\notag
J_{\Lambda_c}(x)&&= \varepsilon_{a'b'c'}u^{a'T}(x)\mathcal{C}\gamma_5 d^{b'}(x)c^{c'}(x), \\
\notag
J_{\Sigma_c}(x)&&= \varepsilon_{a'b'c'}u^{a'T}(x)\mathcal{C}\gamma_\alpha d^{b'}(x)\gamma^\alpha\gamma_5 c^{c'}(x), \\
\notag
J_{\bar D}(y)&&= \bar c^{i'}(y)i\gamma_5 u^{i'}(y), \\
\notag
J_{\bar D^*,\mu}(y)&&= \bar c^{i'}(y)\gamma_\mu u^{i'}(y), \\
J_p(y)&&= \varepsilon_{a'b'c'}u^{a'T}(y)\mathcal{C}\gamma_\alpha u^{b'}(y)\gamma^\alpha\gamma_5 d^{c'}(y), 
\end{eqnarray}
where $a'$, $b'$ $c'$ and $i'$ are color indices.

At the hadron side, we insert complete sets of intermediate hadron states with the same quantum numbers as the corresponding interpolating currents into the three-point correlation function. After performing the integrals in coordinate space, isolating the contributions of the ground states and using the dispersion relation~\cite{Wang:2019hyc,Wang:2020rdh}, we obtain
\begin{eqnarray}\label{eq:4}
\notag
\Pi^{i\mathrm{had}}_{\eta_c p}(p,p')&&=\frac{\langle 0|J_{\eta_c}(0)|\eta_c(p)\rangle\langle 0|J_p(0)|p(q)\rangle}{(m_{\eta_c}^2-p^2)(m_p^2-q^2)(m_{P_{ci}}^2-p'^2)}\\
\notag
&&\times\langle \eta_c(p)p(q)|P_{ci}(p')\rangle\langle P_{ci}(p')|\bar J_{i}(0)|0\rangle+h.c. ,\\
\notag
\Pi^{i\mathrm{had}}_{J/\psi p,\mu}(p,p')&&=\frac{\langle 0|J_{J/\psi,\mu}(0)|J/\psi(p)\rangle\langle 0|J_p(0)|p(q)\rangle}{(m_{J/\psi}^2-p^2)(m_p^2-q^2)(m_{P_{ci}}^2-p'^2)}\\
\notag
&&\times\langle J/\psi(p)p(q)|P_{ci}(p')\rangle\langle P_{ci}(p')|\bar J_{P_{ci}}(0)|0\rangle+h.c. ,\\
\notag
\Pi^{i\mathrm{had}}_{\Lambda_c\bar D}(p,p')&&=\frac{\langle 0|J_{\Lambda_c}(0)|\Lambda_c(p)\rangle\langle 0|J_{\bar D}(0)|\bar D(q)\rangle}{(m_{\Lambda_c}^2-p^2)(m_D^2-q^2)(m_{P_{ci}}^2-p'^2)}\\
\notag
&&\times\langle \Lambda_c(p)\bar D(q)|P_{ci}(p')\rangle\langle P_{ci}(p')|\bar J_{P_{ci}}(0)|0\rangle+h.c. ,\\
\notag
\Pi^{i\mathrm{had}}_{\Lambda_c\bar D^*,\mu}(p,p')&&=\frac{\langle 0|J_{\Lambda_c}(0)|\Lambda_c(p)\rangle\langle 0|J_{\bar D^*,\mu}(0)|\bar D^*(q)\rangle}{(m_{\Lambda_c}^2-p^2)(m_{\bar D^*}^2-q^2)(m_{P_{ci}}^2-p'^2)}\\
\notag
&&\times\langle \Lambda_c(p)\bar D^*(q)|P_{ci}(p')\rangle\langle P_{ci}(p')|\bar J_{P_{ci}}(0)|0\rangle+h.c. ,\\
\notag
\Pi^{i\mathrm{had}}_{\Sigma_c\bar D}(p,p')&&=\frac{\langle 0|J_{\Sigma_c}(0)|\Sigma_c(p)\rangle\langle 0|J_{\bar D}(0)|\bar D(q)\rangle}{(m_{\Sigma_c}^2-p^2)(m_D^2-q^2)(m_{P_{ci}}^2-p'^2)}\\
\notag
&&\times\langle \Sigma_c(p)\bar D(q)|P_{ci}(p')\rangle\langle P_{ci}(p')|\bar J_{P_{ci}}(0)|0\rangle+h.c..\\
\end{eqnarray}
Here $p'=p+q$ and $h.c.$ denote the contributions for higher resonances and continuum states of hadrons. The hadron vacuum matrix elements can be parameterized as following pole residues and decay constants:
\begin{eqnarray}\label{eq:5}
\notag
\langle 0|J_p(0)|p(q)\rangle&&=\lambda_p U_p(q,m_p) ,\\
\notag
\langle 0|J_{\Lambda_c}(0)|\Lambda_c(p)\rangle&&=\lambda_{\Lambda_c} U_{\Lambda_c}(p,m_{\Lambda_c}) ,\\
\notag
\langle 0|J_{\Sigma_c}(0)|\Sigma_c(p)\rangle&&=\lambda_{\Sigma_c} U_{\Sigma_c}(p,m_{\Sigma_c}) ,\\
\notag
\langle 0|J_{\eta_c}(0)|\eta_c(p)\rangle&&=\frac{f_{\eta_c}m_{\eta_c}^2}{2m_c} ,\\
\notag
\langle 0|J_{J/\psi,\mu}(0)|J/\psi(p)\rangle&&=f_{J/\psi}m_{J/\psi}\epsilon^{J/\psi}_\mu(p) ,\\
\notag
\langle 0|J_{\bar{D}}(0)|\bar{D}(q)\rangle&&=\frac{f_{\bar{D}}m_{\bar{D}}^2}{m_c} ,\\
\notag
\langle 0|J_{\bar D^{*},\mu}(0)|\bar D^{*}(q)\rangle&&=f_{\bar{D}^{*}}m_{\bar{D}^{*}}\epsilon^{\bar{D}^*}_\mu(q) ,\\
\langle P_{ci}(p')|\bar J_{P_{ci}}(0)|0\rangle&&=\lambda_{P_{ci}} \bar U_{P_{ci}}(p',m_{P_{ci}}).
\end{eqnarray}
where $U_p$, $U_{\Lambda_c}$, $U_{\Sigma_c}$ and $U_{P_{ci}}$ are the spinor wave functions of corresponding hadrons, and $\epsilon_\mu$ denotes the polarization vector of the vector mesons $J/\psi$ and $\bar D^*$. The strong vertex matrix elements can be defined as the following coupling constants,
\begin{eqnarray}\label{eq:6}
\notag
\langle p(q)\eta_c(p)|P_{ci}(p')\rangle&=&iG_{P_{ci}\eta_c p}\bar U_p(q,m_p)U_{P_{ci}}(p',m_{P_{ci}}),\\
\notag
\langle p(q)J/\psi(p)|P_{ci}(p')\rangle&=&\bar U_p(q,m_{p})\epsilon^{J/\psi*}_\alpha(p)\Bigg[f_{P_{ci}J/\psi p}\gamma^\alpha\\
\notag
&&-\frac{i g_{P_{ci}J/\psi p}}{m_{P_{ci}}+m_p}\sigma^{\alpha\beta}p_\beta\Bigg]\gamma_5 U_{P_{ci}}(p',m_{P_{ci}}) ,\\
\notag
\langle \Lambda_c(p)\bar D(q)|P_{ci}(p')\rangle&=&iG_{P_{ci}\bar D\Lambda_c}\bar U_{\Lambda_c}(p,m_{\Lambda_c})U_{P_{ci}}(p',m_{P_{ci}}) ,\\
\notag
\langle \Lambda_c(p)\bar D^{*}(q)|P_{ci}(p')\rangle&=&\bar U_{\Lambda_c}(p,m_{\Lambda_c})\epsilon^{\bar{D}^{*}*}_\alpha(q)\Bigg[f_{P_{ci}\bar D^{*}\Lambda_c}\gamma^\alpha \\
\notag
&&-\frac{i g_{P_{ci}\bar D^{*}\Lambda_c}}{m_{P_{ci}}+m_{\Lambda_c}}\sigma^{\alpha\beta}q_\beta\Bigg]\gamma_5 U_{P_{ci}}(p',m_{P_{ci}}) ,\\
\notag
\langle \Sigma_c(p)\bar D(q)|P_{ci}(p')\rangle&=&iG_{P_{ci}\bar D\Sigma_c}\bar U_{\Sigma_c}(p,m_{\Sigma_c})U_{P_{ci}}(p',m_{P_{ci}}) .\\
\end{eqnarray}
Here, $G$, $f$ and $g$ are strong couple constants. Substituting the matrix elements in Eq.~(\ref{eq:4}) with Eqs. (\ref{eq:5}) and (\ref{eq:6}), the correlation functions at hadron side can be written as:

\begin{eqnarray}\label{eq:7}
\Pi^{i\mathrm{had}}_{\eta_c p}&=&\frac{i\lambda_p\lambda_{P_{ci}}f_{\eta_c}m_{\eta_c}^2G}{2m_c(m_{\eta_c}^2-p^2)}\frac{(\slashed{q}+m_p)(\slashed{p}'+m_{P_{ci}})}{(m_p^2-q^2)(m_{P_{ci}}^2-p'^2)}+h.c.,
\notag\\
\Pi^{i\mathrm{had}}_{J/\psi p,\mu}&=&\frac{\lambda_p\lambda_{P_{ci}}f_{J/\psi}m_{J/\psi}}{m_{J/\psi}^2-p^2}(\slashed{q}+m_p)\left[f\gamma^\nu-\frac{ig\sigma^{\nu\sigma}p_\sigma}{m_{P_{ci}}+m_p}\right]\gamma_5
\notag\\
&&\times\frac{\slashed{p}+\slashed{q}+m_{P_{ci}}}{(m_p^2-q^2)(m_{P_{ci}}^2-p'^2)}\left(-g_{\mu\nu}+\frac{p_\mu p_\nu}{m_{J/\psi}^2}\right)+h.c.,
\notag\\
\Pi^{i\mathrm{had}}_{\Lambda_c\bar D}&=&\frac{i\lambda_{\Lambda_c}\lambda_{P_{ci}}f_{\bar{D}}m_{\bar{D}}^2G}{m_c(m_{\Lambda_c}^2-p^2)}\frac{(\slashed{p}+m_{\Lambda_c})(\slashed{p}'+m_{P_{ci}})}{(m_{\bar{D}}^2-q^2)(m_{P_{ci}}^2-p'^2)}+h.c.,
\notag\\
\Pi^{i\mathrm{had}}_{\Lambda_c\bar D^*,\mu}&=&\frac{\lambda_{\Lambda_c}\lambda_{P_{ci}}f_{\bar D^*}m_{\bar D^*}}{m_{\Lambda_c}^2-p^2}(\slashed{p}+m_{\Lambda_c})\left[f\gamma^\nu-\frac{ig\sigma^{\nu\sigma}q_\sigma}{m_{P_{ci}}+m_{\Lambda_c}}\right]\gamma_5
\notag\\
&&\times\frac{\slashed{p}+\slashed{q}+m_{P_{ci}}}{(m_{\bar D^*}^2-q^2)(m_{P_{ci}}^2-p'^2)}\left(-g_{\mu\nu}+\frac{q_\mu q_\nu}{m_{\bar D^*}^2}\right)+h.c.,
\notag\\
\Pi^{i\mathrm{had}}_{\Sigma_c\bar D}&=&\frac{i\lambda_{\Sigma_c}\lambda_{P_{ci}}f_{\bar{D}}m_{\bar{D}}^2G}{m_c(m_{\Sigma_c}^2-p^2)}\frac{(\slashed{p}+m_{\Sigma_c})(\slashed{p}'+m_{P_{ci}})}{(m_{\bar{D}}^2-q^2)(m_{P_{ci}}^2-p'^2)}+h.c. .
\end{eqnarray}

The QCD sum rules are obtained by matching the selected structures of the hadron representation with those of the QCD representation. Equivalently, after multiplying the correlation functions by a suitable matrix $\Gamma$ in the Dirac space and taking the trace, we match the invariant amplitudes of the same Lorentz structures at both hadron and QCD side. For the pseudoscalar meson channels, we take $\Gamma=\mathbf{1}$, and for the vector meson channels, we choose $\Gamma=\gamma_5\slashed{z}$ and project out the structures $q_\mu p\cdot z$ and $q_\mu q\cdot z$, where $z_\mu$ is an auxiliary four-vector. The selected structures are written as:

\begin{eqnarray}\label{eq:8}
\frac{1}{4}\mathrm{Tr}\left[\Pi^{i\mathrm{had}}_{\eta_c p}(p,p')\,\mathbf{1}\right]&=&\hat{\Pi}^{i\mathrm{had}}_{\eta_c p}(p'^2,p^2,q^2),
\notag\\
\frac{1}{4}\mathrm{Tr}\left[\Pi^{i\mathrm{had}}_{J/\psi p,\mu}(p,p')\gamma_5\slashed{z}\right]&=&\hat{\Pi}^{i\mathrm{had}(a)}_{J/\psi p}(p^{\prime 2},p^2,q^2)\,q_\mu\, p\cdot z
\notag\\
&&+\hat{\Pi}^{i\mathrm{had}(b)}_{J/\psi p}(p^{\prime 2},p^2,q^2)\,q_\mu\, q\cdot z
\notag\\&&+\cdots ,
\notag\\
\frac{1}{4}\mathrm{Tr}\left[\Pi^{i\mathrm{had}}_{\Lambda_c\bar D}(p,p')\,\mathbf{1}\right]&=&\hat{\Pi}^{i\mathrm{had}}_{\Lambda_c\bar D}(p^{\prime 2},p^2,q^2) ,
\notag\\
\frac{1}{4}\mathrm{Tr}\left[\hat{\Pi}^{i\mathrm{had}}_{\Lambda_c\bar D^*,\mu}(p,p')\gamma_5\slashed{z}\right]&=&\hat{\Pi}^{i\mathrm{had}(a)}_{\Lambda_c\bar D^*}(p^{\prime 2},p^2,q^2)\,q_\mu\, p\cdot z
\notag\\
&&+\hat{\Pi}^{i\mathrm{had}(b)}_{\Lambda_c\bar D^*}(p^{\prime 2},p^2,q^2)\,q_\mu\, q\cdot z
\notag\\
&&+\cdots ,
\notag\\
\frac{1}{4}\mathrm{Tr}\left[\Pi^{i\mathrm{had}}_{\Sigma_c\bar D}(p,p')\,\mathbf{1}\right]&=&\hat{\Pi}^{i\mathrm{had}}_{\Sigma_c\bar D}(p^{\prime 2},p^2,q^2) .
\end{eqnarray}
where $\hat{\Pi}$ is commonly called as the scalar invariant amplitude.

At the QCD side, we substitute the interpolating
currents from Eqs.~(\ref{eq:1}) and (\ref{eq:3}) into Eq.~(\ref{eq:2}) and contract the quark fields with Wick's theorem and express the correlation functions in terms of the full quark propagators. After these steps, the correlation functions at QCD side can be expressed as the following forms,
\begin{widetext}
\begin{eqnarray}\label{eq:9}
	\notag
	\Pi^{1\mathrm{QCD}}_{\eta_c p}(p,p')&=&i^2\frac{2\varepsilon_{ijk}\varepsilon_{mnl}\varepsilon_{kla}\varepsilon_{a'b'c'}}{\sqrt 2}\int d^4x\,d^4y\,e^{ip\cdot x}e^{iq\cdot y}\gamma_\alpha\gamma_5D^{c'j}(y)\gamma_5\mathcal{C} U^{a'iT}(y)\mathcal{C}\gamma_\alpha U^{b'm}(y)\gamma_5\mathcal{C} C^{i'nT}(x)\mathcal{C} i\gamma_5 \mathcal{C} C^{ai'T}(-x)\mathcal{C},\\
	\notag
	\Pi^{1\mathrm{QCD}}_{J/\psi p,\mu}(p,p')&=&-i^2\frac{2\varepsilon_{ijk}\varepsilon_{mnl}\varepsilon_{kla}\varepsilon_{a'b'c'}}{\sqrt 2}\int d^4x\,d^4y\,e^{ip\cdot x}e^{iq\cdot y}\gamma_\alpha\gamma_5D^{c'j}(y)\gamma_5\mathcal{C} U^{a'iT}(y)\mathcal{C}\gamma_\alpha U^{b'm}(y)\gamma_5\mathcal{C} C^{i'nT}(x)\mathcal{C} \gamma_\mu \mathcal{C} C^{ai'T}(-x)\mathcal{C},\\
	\notag
	\Pi^{1\mathrm{QCD}}_{\bar D\Lambda_c}(p,p')&=&-i^2\frac{\varepsilon_{ijk}\varepsilon_{mnl}\varepsilon_{kla}\varepsilon_{a'b'c'}}{\sqrt 2}\int d^4x\,d^4y\,e^{ip\cdot x}e^{iq\cdot y}\Big\{\mathrm{Tr}\big[\mathcal{C} U^{a'iT}(x)\mathcal{C}\gamma_5D^{b'j}(x)\gamma_5\big]C^{c'n}(x)\gamma_5 \mathcal{C} U^{i'mT}(y)\mathcal{C} i\gamma_5 \mathcal{C} C^{ai'T}(-y)\mathcal{C} \\
	\notag
	&&-C^{c'n}(x)\gamma_5 \mathcal{C} U^{a'mT}(x)\mathcal{C}\gamma_5D^{b'j}(x)\gamma_5\mathcal{C} U^{i'iT}(y)\mathcal{C} i\gamma_5 \mathcal{C} C^{ai'T}(-y)\mathcal{C}\Big\},\\
	\notag
	\Pi^{1\mathrm{QCD}}_{\bar D^*\Lambda_c,\mu}(p,p')&=&i^2\frac{\varepsilon_{ijk}\varepsilon_{mnl}\varepsilon_{kla}\varepsilon_{a'b'c'}}{\sqrt 2}\int d^4x\,d^4y\,e^{ip\cdot x}e^{iq\cdot y}\Big\{\mathrm{Tr}\big[\mathcal{C} U^{a'iT}(x)\mathcal{C}\gamma_5D^{b'j}(x)\gamma_5\big]C^{c'n}(x)\gamma_5 \mathcal{C} U^{i'mT}(y)\mathcal{C}\gamma_\mu \mathcal{C} C^{ai'T}(-y)\mathcal{C} \\
	&&-C^{c'n}(x)\gamma_5 \mathcal{C} U^{a'mT}(x)\mathcal{C}\gamma_5D^{b'j}(x)\gamma_5\mathcal{C} U^{i'iT}(y)\mathcal{C}\gamma_\mu \mathcal{C} C^{ai'T}(-y)\mathcal{C}\Big\}.
\end{eqnarray}
\begin{eqnarray}\label{eq:10}
	\notag
	\Pi^{2\mathrm{QCD}}_{\eta_c p}(p,p')&=&i^2\frac{2\varepsilon_{ijk}\varepsilon_{mnl}\varepsilon_{kla}\varepsilon_{a'b'c'}}{\sqrt 2}\int d^4x\,d^4y\,e^{ip\cdot x}e^{iq\cdot y}\gamma_\alpha\gamma_5D^{c'j}(y)\gamma_5\mathcal{C} U^{a'iT}(y)\mathcal{C}\gamma_\alpha U^{b'm}(y)\gamma_\mu \mathcal{C} C^{i'nT}(x)\mathcal{C} i\gamma_5 \mathcal{C} C^{ai'T}(-x)\mathcal{C}\gamma^\mu\gamma_5,\\
	\notag
	\Pi^{2\mathrm{QCD}}_{J/\psi p,\nu}(p,p')&=&-i^2\frac{2\varepsilon_{ijk}\varepsilon_{mnl}\varepsilon_{kla}\varepsilon_{a'b'c'}}{\sqrt 2}\int d^4x\,d^4y\,e^{ip\cdot x}e^{iq\cdot y}\gamma_\alpha\gamma_5D^{c'j}(y)\gamma_5\mathcal{C} U^{a'iT}(y)\mathcal{C}\gamma_\alpha U^{b'm}(y)\gamma_\mu \mathcal{C} C^{i'nT}(x)\mathcal{C}\gamma_\nu \mathcal{C} C^{ai'T}(-x)\mathcal{C}\gamma^\mu\gamma_5,\\
	\notag
	\Pi^{2\mathrm{QCD}}_{\bar D\Lambda_c}(p,p')&=&-i^2\frac{\varepsilon_{ijk}\varepsilon_{mnl}\varepsilon_{kla}\varepsilon_{a'b'c'}}{\sqrt 2}\int d^4x\,d^4y\,e^{ip\cdot x}e^{iq\cdot y}\Big\{\mathrm{Tr}\big[\mathcal{C} U^{a'iT}(x)\mathcal{C}\gamma_5D^{b'j}(x)\gamma_5\big]C^{c'n}(x)\gamma_\mu \mathcal{C} U^{i'mT}(y)\mathcal{C} i\gamma_5 \mathcal{C} C^{ai'T}(-y)\mathcal{C}\gamma^\mu\gamma_5 \\
	\notag
	&&-C^{c'n}(x)\gamma_\mu \mathcal{C} U^{a'mT}(x)\mathcal{C}\gamma_5D^{b'j}(x)\gamma_5\mathcal{C} U^{i'iT}(y)\mathcal{C} i\gamma_5 \mathcal{C} C^{ai'T}(-y)\mathcal{C}\gamma^\mu\gamma_5\Big\},\\
	\notag
	\Pi^{2\mathrm{QCD}}_{\bar D\Sigma_c}(p,p')&=&-i^2\frac{\varepsilon_{ijk}\varepsilon_{mnl}\varepsilon_{kla}\varepsilon_{a'b'c'}}{\sqrt 2}\int d^4x\,d^4y\,e^{ip\cdot x}e^{iq\cdot y}\Big\{\mathrm{Tr}\big[\mathcal{C} U^{a'iT}(x)\mathcal{C}\gamma_\alpha D^{b'j}(x)\gamma_5\big]\gamma^\alpha\gamma_5 C^{c'n}(x)\gamma_\mu \mathcal{C} U^{i'mT}(y)\mathcal{C} i\gamma_5 \mathcal{C} C^{ai'T}(-y)\mathcal{C}\gamma^\mu\gamma_5 \\
	\notag
	&&-\gamma^\alpha\gamma_5 C^{c'n}(x)\gamma_\mu \mathcal{C} U^{a'mT}(x)\mathcal{C}\gamma_\alpha D^{b'j}(x)\gamma_5\mathcal{C} U^{i'iT}(y)\mathcal{C} i\gamma_5 \mathcal{C} C^{ai'T}(-y)\mathcal{C}\gamma^\mu\gamma_5\Big\},\\
	\notag
	\Pi^{2\mathrm{QCD}}_{\bar D^*\Lambda_c,\nu}(p,p')&=&i^2\frac{\varepsilon_{ijk}\varepsilon_{mnl}\varepsilon_{kla}\varepsilon_{a'b'c'}}{\sqrt 2}\int d^4x\,d^4y\,e^{ip\cdot x}e^{iq\cdot y}\Big\{\mathrm{Tr}\big[\mathcal{C} U^{a'iT}(x)\mathcal{C}\gamma_5D^{b'j}(x)\gamma_5\big]C^{c'n}(x)\gamma_\mu \mathcal{C} U^{i'mT}(y)\mathcal{C}\gamma_\nu \mathcal{C} C^{ai'T}(-y)\mathcal{C}\gamma^\mu\gamma_5 \\
	&&-C^{c'n}(x)\gamma_\mu \mathcal{C} U^{a'mT}(x)\mathcal{C}\gamma_5D^{b'j}(x)\gamma_5\mathcal{C} U^{i'iT}(y)\mathcal{C}\gamma_\nu \mathcal{C} C^{ai'T}(-y)\mathcal{C}\gamma^\mu\gamma_5\Big\}.
\end{eqnarray}
\begin{eqnarray}\label{eq:11}
	\notag
	\Pi^{3\mathrm{QCD}}_{\eta_c p}(p,p')&=&-i^2\frac{2\varepsilon_{ijk}\varepsilon_{mnl}\varepsilon_{kla}\varepsilon_{a'b'c'}}{\sqrt 2}\int d^4x\,d^4y\,e^{ip\cdot x}e^{iq\cdot y}\Big\{\mathrm{Tr}\big[\mathcal{C} U^{a'iT}(y)\mathcal{C}\gamma_\sigma U^{b'j}(y)\gamma_\mu\big]\gamma_\sigma\gamma_5D^{c'm}(y)\gamma^\mu \mathcal{C} C^{i'nT}(x)\mathcal{C} i\gamma_5 \mathcal{C} C^{ai'T}(-x)\mathcal{C} \\
	\notag
	&&-\gamma_\sigma\gamma_5D^{a'i}(y)\gamma_\mu \mathcal{C} U^{b'jT}(y)\mathcal{C}\gamma_\sigma U^{c'm}(y)\gamma^\mu \mathcal{C} C^{i'nT}(x)\mathcal{C} i\gamma_5 \mathcal{C} C^{ai'T}(-x)\mathcal{C}\Big\}, \\
	\notag
	\Pi^{3\mathrm{QCD}}_{J/\psi p,\nu}(p,p')&=&-i^2\frac{2\varepsilon_{ijk}\varepsilon_{mnl}\varepsilon_{kla}\varepsilon_{a'b'c'}}{\sqrt 2}\int d^4x\,d^4y\,e^{ip\cdot x}e^{iq\cdot y}\Big\{\mathrm{Tr}\big[\mathcal{C} U^{a'iT}(y)\mathcal{C}\gamma_\sigma U^{b'j}(y)\gamma_\mu\big]\gamma_\sigma\gamma_5D^{c'm}(y)\gamma^\mu \mathcal{C} C^{i'nT}(x)\mathcal{C}\gamma_\nu \mathcal{C} C^{ai'T}(-x)\mathcal{C} \\
	\notag
	&&-\gamma_\sigma\gamma_5D^{a'i}(y)\gamma_\mu \mathcal{C} U^{b'jT}(y)\mathcal{C}\gamma_\sigma U^{c'm}(y)\gamma^\mu \mathcal{C} C^{i'nT}(x)\mathcal{C}\gamma_\nu \mathcal{C} C^{ai'T}(-x)\mathcal{C}\Big\},\\
	\notag
	\Pi^{3\mathrm{QCD}}_{\bar D\Lambda_c}(p,p')&=&-i^2\frac{\varepsilon_{ijk}\varepsilon_{mnl}\varepsilon_{kla}\varepsilon_{a'b'c'}}{\sqrt 2}\int d^4x\,d^4y\,e^{ip\cdot x}e^{iq\cdot y}\Big\{2\mathcal{C}C^{ai'}(-y)i\gamma_5 U^{i'j}(y)\gamma_\mu \mathcal{C} U^{a'iT}(x)\mathcal{C}\gamma_5D^{b'm}(x)\gamma^\mu \mathcal{C} C^{c'nT}(x) \\
	\notag
	&&-\mathrm{Tr}\big[\mathcal{C} U^{a'iT}(x)\mathcal{C}\gamma_5D^{b'j}(x)\gamma_\mu\big]C^{c'n}(x)\gamma^\mu \mathcal{C} U^{i'mT}(y)\mathcal{C} i\gamma_5 \mathcal{C} C^{ai'T}(-y)\mathcal{C} \\
	&&+C^{c'n}(x)\gamma^\mu \mathcal{C} U^{a'mT}(x)\mathcal{C}\gamma_5D^{b'j}(x)\gamma_\mu \mathcal{C} U^{i'iT}(y)\mathcal{C} i\gamma_5 \mathcal{C} C^{ai'T}(-y)\mathcal{C}\Big\}.
\end{eqnarray}
\begin{eqnarray}\label{eq:12}
	\notag
	\Pi^{4\mathrm{QCD}}_{\eta_c p}(p,p')&=&-i^2\frac{2\varepsilon_{ijk}\varepsilon_{mnl}\varepsilon_{kla}\varepsilon_{a'b'c'}}{\sqrt 2}\int d^4x\,d^4y\,e^{ip\cdot x}e^{iq\cdot y}\Big\{\mathrm{Tr}\big[\mathcal{C} U^{a'iT}(y)\mathcal{C}\gamma_\alpha U^{b'j}(y)\gamma_\mu
	\big]\gamma_\alpha\gamma_5D^{c'm}(y)\gamma_5\mathcal{C} C^{i'nT}(x)\mathcal{C} i\gamma_5 \mathcal{C} C^{ai'T}(-x)\mathcal{C}\gamma^\mu\gamma_5 \\
	\notag
	&&-\gamma_\alpha\gamma_5D^{c'j}(y)\gamma_\mu \mathcal{C} U^{a'iT}(y)\mathcal{C}\gamma_\alpha U^{b'm}(y)\gamma_5\mathcal{C} C^{i'nT}(x)\mathcal{C} i\gamma_5 \mathcal{C} C^{ai'T}(-x)\mathcal{C}\gamma^\mu\gamma_5\Big\},\\
	\notag
	\Pi^{4\mathrm{QCD}}_{J/\psi p,\nu}(p,p')&=&-i^2\frac{2\varepsilon_{ijk}\varepsilon_{mnl}\varepsilon_{kla}\varepsilon_{a'b'c'}}{\sqrt 2}\int d^4x\,d^4y\,e^{ip\cdot x}e^{iq\cdot y}\Big\{\mathrm{Tr}\big[\mathcal{C} U^{a'iT}(y)\mathcal{C}\gamma_\alpha U^{b'j}(y)\gamma_\mu\big]\gamma_\alpha\gamma_5D^{c'm}(y)\gamma_5\mathcal{C} C^{i'nT}(x)\mathcal{C}\gamma_\nu \mathcal{C} C^{ai'T}(-x)\mathcal{C}\gamma^\mu\gamma_5 \\
	\notag
	&&-\gamma_\alpha\gamma_5D^{c'j}(y)\gamma_\mu \mathcal{C} U^{a'iT}(y)\mathcal{C}\gamma_\alpha U^{b'm}(y)\gamma_5\mathcal{C} C^{i'nT}(x)\mathcal{C}\gamma_\nu \mathcal{C} C^{ai'T}(-x)\mathcal{C}\gamma^\mu\gamma_5\Big\}, \\
	\notag
	\Pi^{4\mathrm{QCD}}_{\bar D\Lambda_c}(p,p')&=&-i^2\frac{\varepsilon_{ijk}\varepsilon_{mnl}\varepsilon_{kla}\varepsilon_{a'b'c'}}{\sqrt 2}\int d^4x\,d^4y\,e^{ip\cdot x}e^{iq\cdot y}\big\{2\mathcal{C}\gamma_5\gamma^\mu C^{ai'}(-y)i\gamma_5 U^{i'j}(y)\gamma_\mu \mathcal{C} U^{a'iT}(x)\mathcal{C}\gamma_5D^{b'm}(x)\gamma_5 \mathcal{C} C^{c'nT}(x) \\
	\notag
	&&-\mathrm{Tr}\big[\mathcal{C} U^{a'iT}(x)\mathcal{C}\gamma_5D^{b'j}(x)\gamma_\mu\big]C^{c'n}(x)\gamma_5 \mathcal{C} U^{i'mT}(y)\mathcal{C} i\gamma_5 \mathcal{C} C^{ai'T}(-y)\mathcal{C}\gamma^\mu\gamma_5 \\
	&&+C^{c'n}(x)\gamma_5 \mathcal{C} U^{a'mT}(x)\mathcal{C}\gamma_5D^{b'j}(x)\gamma_\mu \mathcal{C} U^{i'iT}(y)\mathcal{C} i\gamma_5 \mathcal{C} C^{ai'T}(-y)\mathcal{C}\gamma^\mu\gamma_5\Big\}.
\end{eqnarray}
\end{widetext}
where $U[D]^{ij}(x)$ and $C^{ij}(x)$ are the full propagator of $u(d)$ and $c$ quarks which can be written as follows~\cite{Reinders:1984sr,Wang:2013vex}: 
\begin{eqnarray}\label{eq:13}
\notag
U[D]^{ij}(x) &&= \frac{i\delta ^{ij}\slashed x}{2\pi ^2x^4} - \frac{\delta ^{ij}\left\langle \bar qq \right\rangle}{12} - \frac{\delta ^{ij}x^2\left\langle \bar qg_s\sigma Gq \right\rangle }{192}\\
\notag
&&- \frac{ig_sG_{\lambda \tau }^at_{ij}^a(\slashed x\sigma ^{\lambda \tau } + \sigma ^{\lambda \tau }\slashed x)}{32\pi ^2x^2} - \frac{\left\langle \bar q^j\sigma ^{\lambda \tau }q^i \right\rangle \sigma _{\lambda \tau }}{8}\\
\notag
&&-\frac{\delta^{ij}x^4\left\langle \bar qq \right\rangle \left\langle g_s^2 GG \right\rangle}{27648}+ ...,
\end{eqnarray}
\begin{eqnarray}
\notag
C^{ij}(x) &&= \frac{i}{(2\pi )^4}\int d^4k e^{ - ikx} \left\{ \frac{\delta ^{ij}(\slashed{k}+m_c)}{k^2 - m_c^2} \right.\\
\notag
&&- \frac{g_sG_{\lambda \tau }^nt_{ij}^n}{4}\frac{\sigma ^{\lambda \tau }(\slashed k + m_c) + (\slashed k + m_c)\sigma ^{\lambda \tau }}{(k^2 - m_c^2)^2}\\
&&\left. { +\frac{\delta ^{ij}\langle g_s^2GG\rangle(m_c k^2+m_c^2\slashed{k})}{12(k^2 - m_c^2)^4} + ...} \right\},
\end{eqnarray}
here, $\langle\bar{q}g_s\sigma Gq\rangle=\langle\bar{q}g_s\sigma_{\lambda\tau} G^{\alpha}_{\lambda\tau}t^{\alpha}q\rangle$, $\langle g_s^2GG\rangle=\langle g_s^2G^c_{\alpha\beta}G^{c\alpha\beta}\rangle$, $t^\alpha=\frac{\lambda^\alpha}{2}$ and $\lambda^\alpha$ ($\alpha=1,...,8$) are the Gell-Mann matrices.

After carrying out the operator product expansion, we project out the selected structures consistent with one side of the hadron and calculate the corresponding scalar invariant amplitudes. In this analysis, the scalar invariant amplitude at QCD side can be expressed as the following form by double dispersion relation:
\begin{eqnarray}\label{eq:14}
\notag
\hat{\Pi}^{i\mathrm{QCD}}(p'^2,p^2,q^2)&=&\int_{\Delta_s^2}^{\infty}ds\int_{\Delta_u^2}^{\infty}du\frac{\rho^{i\mathrm{QCD}}(p'^2,s,u)}{(s-p^2)(u-q^2)},\\
\end{eqnarray}
where $\Delta_s^2$ and $\Delta_u^2$ are kinetic limit for hadrons and their values are usually taken as the square of the summation of the quark masses that make up the corresponding hadrons, $\rho^{i\mathrm{QCD}}$ represent the QCD spectral density with $s=p^2$, $u=q^2$ and can be written as:
\begin{eqnarray}\label{eq:15}
\rho^{i\mathrm{QCD}}(p'^2,s,u)&=&\frac{1}{(2\pi i)^2}\mathrm{Disc}_s\mathrm{Disc}_u\hat{\Pi}^{i\mathrm{QCD}}(p'^2,s,u),
\end{eqnarray}
here, $\mathrm{Disc}_s$ represents the discontinuity of the variable $s$, and can be expressed as:
\begin{eqnarray}\label{eq:16}
\mathrm{Disc}_sf(s)&=&\lim_{\epsilon\to0}[f(s+i\epsilon)-f(s-i\epsilon)].
\end{eqnarray}
 
It is noted that the quark-hadron duality should be implemented with some care in the three-point QCD sum rules for the strong decay vertices. At the hadron side, the invariant amplitudes admit the triple dispersion representation in the variables $p'^2$, $p^2$ and $q^2$,
\begin{eqnarray}\label{eq:17}
\notag
\hat{\Pi}^{i\mathrm{had}}(p'^2,p^2,q^2)&=&\int_{\Delta_{s'}^2}^{\infty}ds'\int_{\Delta_s^2}^{\infty}ds\int_{\Delta_u^2}^{\infty}du \\
\notag
&&\times\frac{\rho^{i\mathrm{had}}(s',s,u)}{(s'-p^{\prime 2})(s-p^2)(u-q^2)} ,\\
\notag
\rho^{i\mathrm{had}}(s',s,u)&=&\frac{1}{(2\pi i)^3}\mathrm{Disc}_{s'}\mathrm{Disc}_s\mathrm{Disc}_u\hat{\Pi}^{i\mathrm{had}}(s',s,u).\\
\end{eqnarray}
However, the correlation functions have only the double dispersion representation in the two final-state channels at the QCD side, as there is no non-zero triple QCD spectral density in the variables $p'^2$, $p^2$ and $q^2$
\begin{eqnarray}\label{eq:18}
{\rm Disc}_{s'}\mathrm{Disc}_s\mathrm{Disc}_u\hat{\Pi}^{i\mathrm{QCD}}(s',s,u)=0 .
\end{eqnarray}
Therefore, the hadron spectral density $\rho^{i\mathrm{had}}(s',s,u)$ can not be matched directly with a triple QCD spectral density at QCD side. Following the rigorous quark-hadron duality prescription for the three-point QCD sum rules~\cite{Wang:2025sic}, we first integrate over the variable $s'$ in the initial pentaquark channel at the hadron side, and then match the resulting expression with the QCD representation below the continuum thresholds in the two final-state channels~\cite{Wang:2017lot,Wang:2016wkj,Wang:2019iaa,Wang:2018qpe},
\begin{eqnarray}\label{eq:19}
&&\int_{\Delta_s^2}^{s_0}ds\int_{\Delta_u^2}^{u_0}du\int_{\Delta_{s'}^2}^{\infty}ds'\,\frac{\rho^{i\mathrm{had}}(s',s,u)}{(s'-p'^2)(s-p^2)(u-q^2)}
\nonumber\\
&&=\int_{\Delta_s^2}^{s_0}ds\int_{\Delta_u^2}^{u_0}du\,\frac{\rho^{i\mathrm{QCD}}(s,u)}{(s-p^2)(u-q^2)} .
\end{eqnarray}
Here and hereafter, we have written $\rho^{i\mathrm{QCD}}(p'^2,s,u)$ as $\rho^{i\mathrm{QCD}}(s,u)$ for simplicity after the relation $p'^2=\xi p^2$ is imposed. For the hidden-charm channels $P_{ci}\to\eta_c p$ and $P_{ci}\to J/\psi p$, we set $\xi=1$, while for the open-charm channels involving the charmed baryons $\Lambda_c$ or $\Sigma_c$, such as $P_{ci}\to\Lambda_c\bar D$, $P_{ci}\to\Lambda_c\bar D^*$ and $P_{ci}\to\Sigma_c\bar D$, we set $\xi=2$~\cite{Wang:2019hyc,Wang:2020rdh,Wang:2023kir}.

Then, we rewrite the Eq.~(\ref{eq:19}) as the following form,
\begin{eqnarray}\label{eq:20}
\notag
&&\int_{\Delta_s^2}^{s_0}ds\int_{\Delta_u^2}^{u_0}du\int_{\Delta_{s'}^2}^{s'_0}ds'\,\frac{\rho^{i\mathrm{had}}(s',s,u)}{(s'-p'^2)(s-p^2)(u-q^2)}\\
\notag
&&+\int_{\Delta_s^2}^{s_0}ds\int_{\Delta_u^2}^{u_0}du\frac{1}{(s-p^2)(u-q^2)}\left[\int_{s'_0}^{\infty}ds'\frac{\rho^{i\mathrm{had}}(s',s,u)}{s'-p'^2}\right]\\
&&=\int_{\Delta_s^2}^{s_0}ds\int_{\Delta_u^2}^{u_0}du\,\frac{\rho^{i\mathrm{QCD}}(s,u)}{(s-p^2)(u-q^2)} .
\end{eqnarray}
The first term on the left side of Eq.~(\ref{eq:20}) corresponds to the pure ground state contribution represented on the right side of Eq.~(\ref{eq:4}). However, we can not get the specific form of the second term on the left side of Eq.~(\ref{eq:20}).

For illustration, we present explicitly the sum rules for the two independent structures of the $J/\psi p$ channel. By matching the invariant amplitudes of both hadron and QCD sides, the following equations can be obtained,
\begin{eqnarray}\label{eq:21}
\notag
&&\frac{\lambda_p\lambda_{P_{ci}}f_{J/\psi}m_{J/\psi}(f-g)}{(m_{J/\psi}^2-p^2)(m_p^2-q^2)(m_{P_{ci}}^2-p'^2)}+\frac{C_{J/\psi p}^{a}}{(m_{J/\psi}^2-p^2)(m_p^2-q^2)}  \\
\notag
&&=\int_{4m_c^2}^{s_0^{J/\psi}}ds\int_{0}^{u_0^{p}}du\,\frac{\rho^{i\mathrm{QCD}(a)}(s,u)}{(s-p^2)(u-q^2)},\\
\notag
&&\frac{2\lambda_p\lambda_{P_{ci}}f_{J/\psi}m_{J/\psi}f}{(m_{J/\psi}^2-p^2)(m_p^2-q^2)(m_{P_{ci}}^2-p'^2)}+\frac{C_{J/\psi p}^{b}}{(m_{J/\psi}^2-p^2)(m_p^2-q^2)}\\
\notag
&&=\int_{4m_c^2}^{s_0^{J/\psi}}ds\int_{0}^{u_0^{p}}du\,\frac{\rho^{i\mathrm{QCD}(b)}(s,u)}{(s-p^2)(u-q^2)},\\
\end{eqnarray}
here,
\begin{eqnarray}
\notag
C_{J/\psi p}^{a}&=&\int_{s'_0}^{\infty}ds'\frac{\rho^{i\mathrm{had}(a)}(s',m_{J/\psi}^2,m_p^2)}{s'-p'^2}\\
C_{J/\psi p}^{b}&=&\int_{s'_0}^{\infty}ds'\frac{\rho^{i\mathrm{had}(b)}(s',m_{J/\psi}^2,m_p^2)}{s'-p'^2}
\end{eqnarray}
denote the effective subtraction terms~\cite{Wang:2025sic} for the two independent structures of the $J/\psi p$ channel. The effective subtraction parameters $C$ are not new physical strong coupling constants. They are auxiliary subtraction constants introduced at the hadron side to parameterize the unknown pole-continuum transition contributions.

After using the approximate relation $p'^2=\xi p^2$ ($\xi=1\sim4$), taking the variables change $p^2\to-P^2$ and $q^2\to-Q^2$, and performing the double Borel transforms~\cite{Reinders:1984sr} for variables $P^2$ and $Q^2$ to the two side of Eq.~(\ref{eq:21}), the following equations can be obtained,
\begin{eqnarray}\label{eq:23}
\notag
&&\lambda_p\lambda_{P_{ci}}f_{J/\psi}m_{J/\psi}(f-g)\frac{\exp\left(-\dfrac{m_{J/\psi}^2}{T_1^2}\right)-\exp\left(-\dfrac{m_{P_{ci}}^2}{\xi T_1^2}\right)}{\xi\left(\dfrac{m_{P_{ci}}^2}{\xi}-m_{J/\psi}^2\right)\exp\left(\dfrac{m_p^2}{T_2^2}\right)}\\
\notag
&&+C_{J/\psi p}^{a}\exp\left(-\dfrac{m_{J/\psi}^2}{T_1^2}-\dfrac{m_p^2}{T_2^2}\right)=\mathcal{R}_a(T_1^2,T_2^2),\\
\notag
&&2\lambda_p\lambda_{P_{ci}}f_{J/\psi}m_{J/\psi}f\frac{\exp\left(-\dfrac{m_{J/\psi}^2}{T_1^2}\right)-\exp\left(-\dfrac{m_{P_{ci}}^2}{\xi T_1^2}\right)}{\xi\left(\dfrac{m_{P_{ci}}^2}{\xi}-m_{J/\psi}^2\right)\exp\left(\dfrac{m_p^2}{T_2^2}\right)}\\
&&+C_{J/\psi p}^{b}\exp\left(-\dfrac{m_{J/\psi}^2}{T_1^2}-\dfrac{m_p^2}{T_2^2}\right)=\mathcal{R}_b(T_1^2,T_2^2),
\end{eqnarray}
with
\begin{eqnarray}\label{eq:24}
\notag
\mathcal{R}_{a[b]}(T_1^2,T_2^2)&&=\int_{4m_c^2}^{s_0^{J/\psi}}ds\int_0^{u_0^p}du\rho^{i\mathrm{QCD}(a)[(b)]}(s,u)\\
&&\times\exp\left(-\frac{s}{T_1^2}-\frac{u}{T_2^2}\right),
\end{eqnarray}
where, $T_1^2$ and $T_2^2$ are Borel parameters corresponding to variables $P^2$ and $Q^2$, respectively. By solving the above two coupled sum rules, the following QCD sum rule equations for strong coupling constants $f$ and $g$ are obtained:
\begin{eqnarray}\label{eq:25}
\notag
f&=&\frac{\xi\left(\dfrac{m_{P_{ci}}^2}{\xi}-m_{J/\psi}^2\right)\exp\left(\dfrac{m_p^2}{T_2^2}\right)}{2\lambda_p\lambda_{P_{ci}}f_{J/\psi}m_{J/\psi}\exp\left(-\dfrac{m_{J/\psi}^2}{T_1^2}-\dfrac{m_{P_{ci}}^2}{\xi T_1^2}\right)}\\
\notag
&&\times\Bigg[\mathcal{R}_b(T_1^2,T_2^2)-C_f\exp\left(-\dfrac{m_{J/\psi}^2}{T_1^2}-\dfrac{m_p^2}{T_2^2}\right)\Bigg],\\
\notag
g&=&\frac{\xi\left(\dfrac{m_{P_{ci}}^2}{\xi}-m_{J/\psi}^2\right)\exp\left(\dfrac{m_p^2}{T_2^2}\right)}{2\lambda_p\lambda_{P_{ci}}f_{J/\psi}m_{J/\psi}\exp\left(-\dfrac{m_{J/\psi}^2}{T_1^2}-\dfrac{m_{P_{ci}}^2}{\xi T_1^2}\right)}\\
\notag
&&\times\Bigg[\mathcal R_b(T_1^2,T_2^2)-2\mathcal R_a(T_1^2,T_2^2)-C_g\exp\left(-\dfrac{m_{J/\psi}^2}{T_1^2}-\dfrac{m_p^2}{T_2^2}\right)\Bigg],\\
\notag
C_f&=&C_{J/\psi p}^{b},\\
C_g&=&C_{J/\psi p}^{b}-2C_{J/\psi p}^{a}.
\end{eqnarray}
For simplicity, we only show the QCD spectral densities for process $P_{c1}\to J/\psi p$, its full expression are as follows:
\begin{widetext}
\begin{eqnarray}\label{eq:26}
\notag
\rho^{1\mathrm{QCD}(a)}(s,u)&=&\int_{x_{-}}^{x_{+}}dx\frac{m_c u^2}{2048\sqrt{2}\pi^6}+\left\langle\bar qg_s\sigma Gq\right\rangle\int_0^1dx\left[\frac{ux}{9216\sqrt{2}\pi^4}-\frac{m_c^2(\xi-1)}{9216\sqrt{2}\pi^4(1-x)}\right]\delta\left(s-\frac{m_c^2}{x(1-x)}\right)\\
\notag
&&+\langle\bar{q}q\rangle^2\int_{x_{-}}^{x_{+}}dx\frac{m_c}{12\sqrt{2}\pi^2}\delta(u)-\left\langle\bar qq\right\rangle\left\langle\bar qg_s\sigma Gq\right\rangle
\int_{x_-}^{x_+}dx\frac{m_c}{24\sqrt{2}\pi^2T_2^2}\delta(u) \\
\notag
&&+\left\langle\bar qq\right\rangle
\left\langle\bar qg_s\sigma Gq\right\rangle\int_0^1dx\frac{m_c}{\sqrt{2}\pi^2}
\Bigg[\frac{m_c^2(2x-1)}{192T_2^2x(1-x)^2}+\frac{m_c^2(2x-1)(x-5)}{1152T_1^2[x(1-x)]^2}
-\frac{7x^3+6x^2-18x+5}{1152x(1-x)^2}\Bigg]\\
\notag
&&\times\delta\left(s-\frac{m_c^2}{x(1-x)}
\right)\delta(u)+\left\langle\bar qg_s\sigma Gq\right\rangle^2\int_0^1dx\frac{m_c}{\sqrt{2}\pi^2}
\Bigg[\frac{7x^3+6x^2-18x+5}{4608T_2^2x(1-x)^2}-\frac{m_c^2(2x-1)(x-5)}{4608T_1^2T_2^2[x(1-x)]^2}\Bigg]\\
\notag
&&\times\delta\left(s-\frac{m_c^2}{x(1-x)}\right)\delta(u)+\left\langle g_s^2GG\right\rangle
\int_0^1dx\frac{m_c}{\sqrt{2}\pi^6}\Bigg[\frac{m_c^2(2x-1)}{4096x(1-x)^2}+\frac{m_c^2u(2x-1)^2}{36864T_1^2[x(1-x)]^2}+\frac{u(2x^2+2x-1)}{36864x(1-x)}\\
\notag
&&-\frac{m_c^8u^2(2x-1)^3}{884736T_1^{10}[x(1-x)]^6}-\frac{m_c^6u^2x(32x^4-86x^3+85x^2-37x+6)}{294912T_1^8[x(1-x)]^6}-\frac{m_c^4u^2(98x^3-198x^2+137x-32)}{294912T_1^6[x(1-x)]^4}\\
\notag
&&-\frac{m_c^2u^2x(31x^4-131x^3+192x^2-121x+28)}{147456T_1^4[x(1-x)]^4}-\frac{u^2x(6x^4+8x^3-39x^2+39x-12)}{147456T_1^2[x(1-x)]^3}-\frac{u^2}{147456T_1^2x(1-x)}\Bigg]\\
\notag
&&\times\delta\left(s-\frac{m_c^2}{x(1-x)}
\right)-\left\langle g_s^2GG\right\rangle\left\langle\bar qq\right\rangle
\int_0^1dx\frac{m_c^2(\xi-1)}{9216\sqrt{2}\pi^4(1-x)}\delta\left(s-\frac{m_c^2}{x(1-x)}\right)\delta(u),
\end{eqnarray}
\begin{eqnarray}
\notag
\rho^{1\mathrm{QCD}(b)}(s,u)&=&
-\left\langle\bar qg_s\sigma Gq\right\rangle
\int_{x_-}^{x_+}dx\frac{x}{2304\sqrt{2}\pi^4}-\left\langle\bar qg_s\sigma Gq\right\rangle
\int_0^1dx\Bigg[\frac{m_c^4(2x-1)(x-2)}{4608\sqrt{2}\pi^4T_1^2x^2(1-x)^3}+\frac{m_c^2(3x-2)}{4608\sqrt{2}\pi^4x(1-x)}\Bigg]\\
\notag
&&\times\delta\left(s-\frac{m_c^2}{x(1-x)}\right)-\left\langle g_s^2GG\right\rangle
\left\langle\bar qq\right\rangle\int_{x_-}^{x_+}dx\frac{x}{2304\sqrt{2}\pi^4}\delta(u)-\left\langle g_s^2GG\right\rangle\left\langle\bar qq\right\rangle
\int_0^1dx\Bigg[\frac{m_c^4(2x-1)(x-2)}{4608\sqrt{2}\pi^4T_1^2x^2(1-x)^3}\\
&&+\frac{m_c^2(3x-2)}{4608\sqrt{2}\pi^4x(1-x)}\Bigg]\delta\left(s-\frac{m_c^2}{x(1-x)}\right)\delta(u),
\end{eqnarray}
\end{widetext}
where $x_{\pm}=\frac{1}{2}\left(1\pm\sqrt{1-\frac{4m_c^2}{s}}\right)$. In this analysis, we take the relation $T_1^2=T_2^2=T^2$ by merging the two Borel parameters to obtain more stable QCD sum rules~\cite{Wang:2019hyc}.
\section{Numerical results and discussions}\label{sec3}

The input parameters used in the numerical analysis are collected in Table~\ref{tab:1}, and the light quark masses are neglected. The $c$ quark mass and vacuum condensates are energy scale dependent and fulfill the following renormalization group equations~\cite{ParticleDataGroup:2024cfk}:
\begin{eqnarray}\label{eq:27}
\notag
\langle\overline{q}q\rangle(\mu)&=&\langle\overline{q}q\rangle(1{\rm GeV})\left[\frac{\alpha_s(1{\rm GeV})}{\alpha_s(\mu)}\right]^{\frac{12}{33-2n_f}},\\
\notag
\langle\overline{q}g_s\sigma Gq\rangle(\mu)&=&\langle\overline{q}g_s\sigma Gq\rangle(1{\rm GeV})\left[\frac{\alpha_s(1{\rm GeV})}{\alpha_s(\mu)}\right]^{\frac{2}{33-2n_f}},\\
\notag
m_c(\mu)&=&m_c(m_c)\left[\frac{\alpha_s(\mu)}{\alpha_s(m_c)}\right]^{\frac{12}{33-2n_f}},\\
\notag
\alpha_s(\mu)&=&\frac{1}{b_0t}\left[1-\frac{b_1}{b_0^2}\frac{\log t}{t}\right.\\
&+& \left.\frac{b_1^2( \log^2 t-\log t-1)+b_0b_2}{b_0^4t^2}\right],
\end{eqnarray}
where $t=\log(\mu^2/\Lambda_{\mathrm{QCD}}^2)$, $b_0=\frac{33-2n_f}{12\pi}$, $b_1=\frac{153-19n_f}{24\pi^2}$ and $b_2=\frac{2857-\frac{5033}{9}n_f+\frac{325}{27}n_f^2}{128\pi^3}$. $\Lambda_{\mathrm{QCD}}=292$ MeV is for quark flavors $n_f=4$ in this work. The minimum subtraction mass of $c$ quark is taken from the Particle Data Group, which is $m_c(m_c)=1.275\pm0.025$~GeV~\cite{ParticleDataGroup:2024cfk}. The typical energy scale $\mu=1$ GeV is adopted in this analysis which is widely used in charmed systems. 
\begin{table}[htbp]
	\begin{ruledtabular}\caption{Input parameters (IP) used in this work. The values of the vacuum condensates are taken at the energy scale $\mu=1$ GeV.}\label{tab:1}
		\begin{tabular}{c c c c}
			IP & Values (GeV) & IP & Values \\
			\hline
			$m_{\eta_c}$ &$2.984$~\cite{ParticleDataGroup:2024cfk}& $f_{\bar{D}}$ & $0.210$ GeV~\cite{Wang:2015mxa} \\
			$m_{J/\psi}$ & $3.097$~\cite{ParticleDataGroup:2024cfk}
			& $f_{\bar{D}^*}$ & $0.236$ GeV~\cite{Wang:2015mxa} \\
			$m_p$ & $0.938$~\cite{ParticleDataGroup:2024cfk}
			& $\lambda_p$ & $0.032$ GeV$^3$~\cite{Ioffe:2005ym} \\
			$m_{\bar D}$ & $1.86$~\cite{ParticleDataGroup:2024cfk}
			& $\lambda_{\Lambda_c}$ & $0.015$ GeV$^3$~\cite{Wang:2020mxk} \\
			$m_{\bar D^*}$ & $2.01$~\cite{ParticleDataGroup:2024cfk}
			& $\lambda_{\Sigma_c}$ & $0.045$ GeV$^3$~\cite{Wang:2010vn} \\
			$m_{\Lambda_c}$ & $2.286$~\cite{ParticleDataGroup:2024cfk}
			& $\lambda_{P_{c1}}$ & $1.40\times10^{-3}$ GeV$^6$~\cite{Wang:2025qtm} \\
			$m_{\Sigma_c}$ & $2.455$~\cite{ParticleDataGroup:2024cfk}
			& $\lambda_{P_{c2}}$ & $3.02\times10^{-3}$ GeV$^6$~\cite{Wang:2025qtm} \\
			$m_{P_{c1}}$ & $4.31$~\cite{Wang:2025qtm}
			& $\lambda_{P_{c3}}$ & $2.24\times10^{-3}$ GeV$^6$~\cite{Wang:2025qtm} \\
			$m_{P_{c2}}$ & $4.45$~\cite{Wang:2025qtm}
			& $\lambda_{P_{c4}}$ & $2.78\times10^{-3}$ GeV$^6$~\cite{Wang:2025qtm} \\
			$m_{P_{c3}}$ & $4.20$~\cite{Wang:2025qtm}
			& $\langle\bar q q\rangle$ & $-(0.24\pm0.01)^3$ GeV$^3$~\cite{Shifman:1978bx,Shifman:1978by,Reinders:1984sr} \\
			$m_{P_{c4}}$ & $4.25$~\cite{Wang:2025qtm}
			& $\langle\bar q g_s\sigma Gq\rangle$ & $m_0^2\langle\bar q q\rangle$~\cite{Shifman:1978bx,Shifman:1978by,Reinders:1984sr} \\
			$f_{\eta_c}$ & $0.387$~\cite{Becirevic:2013bsa}& $m_0^2$ & $(0.8\pm0.1)$ GeV$^2$~\cite{Shifman:1978bx,Shifman:1978by,Reinders:1984sr} \\
			$f_{J/\psi}$ & $0.418$~\cite{Becirevic:2013bsa}& $\left\langle g_s^2GG\right\rangle$ & $0.47\pm0.15$ GeV$^4$~\cite{Narison:2010cg,Narison:2011xe,Narison:2011rn} \\
		\end{tabular}
	\end{ruledtabular}
\end{table}

The continuum threshold parameters are introduced to eliminate the contributions of higher resonances and continuum states. They commonly satisfy the relation $u^0(s^0)=(m_{\mathrm{ground}}+\delta)^2$, where the $m_{\mathrm{ground}}$ denotes the mass of ground state hadron, and $\delta$ is the energy gap between the ground and first excited states, commonly taken as a value of $0.3-0.9$ GeV, which is based on experimental data and QCD sum rule calculation. In the present work, we take $\sqrt{s_{\eta_c}^0}=3.50\pm0.10$ GeV, $\sqrt{s_{J/\psi}^0}=3.60\pm0.10$ GeV and $\sqrt{u_p^0}=1.30\pm0.10$ GeV for the charmonium and proton~\cite{ParticleDataGroup:2024cfk}. And for the charmed mesons and baryons, we take $u^0_{\bar{D}}=6.20\pm0.5$ GeV$^2$, $u^0_{\bar{D}^*}=6.30\pm0.5$ GeV$^2$~\cite{Wang:2015mxa}, $\sqrt{s^0_{\Lambda_c}}=2.75\pm0.10$ GeV~\cite{Wang:2020mxk} and $\sqrt{s^0_{\Sigma_c}}=3.20\pm0.10$ GeV~\cite{Wang:2010vn}. As a free parameter, the effective subtraction terms $C$ are fitted to obtain flat platforms, which are presented explicitly in Table~\ref{tab:2}. Then we obtain uniform flat platforms $T^2_{\max}-T^2_{\min}=1$ GeV$^2$ for all the decay channels, just like what have been done in our previous works~\cite{Wang:2019hyc,Wang:2020rdh,Wang:2023kir}. The strong coupling constants extracted at the central values of the Borel parameters are also summarized in Table~\ref{tab:2}. The uncertainties of the strong coupling constants originate from the input parameters. These input parameters can be divided into two types: one at hardon level, such as the decay constants and pole residues of hadrons, and the other at QCD level, such as the vacuum condensates, $c$ quark mass and threshold parameters. In the present work, we neglect the uncertainty caused by the mass of hadrons. To estimate the uncertainty of strong coupling constant, we uniformly write all QCD sum rule equations in the following form: 
\begin{eqnarray}\label{eq:28}
f_M\lambda_{P_c}\lambda_{B}G=f(x_i),
\end{eqnarray}
where $f_M$ is the decay constant of meson, $\lambda_{P_c}$ and $\lambda_{B}$ are the pole residues of hidden charm pentaquark and baryon and $G$ denotes the strong coupling constant. $f(x_i)$ is a numerical integral associated with QCD level physical quantities, with $x_i$ denotes the input parameters,
\begin{eqnarray}
\notag
x_i&=&\left\{\langle\bar q q\rangle, \langle\bar q g_s\sigma Gq\rangle, \left\langle g_s^2GG\right\rangle, m_c, s_{\eta_c}^0, s_{J/\psi}^0, u_p^0, s_{\Lambda_c}^0, s_{\Sigma_c}^0,\right. \\
\notag
&&\left. u_D^0, u_{D^*}^0\right\}.
\end{eqnarray} 
The left and right sides of Eq.~(\ref{eq:28}) are quantities at the hadron and QCD level, respectively. After taking the partial derivative of both sides of Eq.~(\ref{eq:28}), we get
\begin{eqnarray}\label{eq:29}
	\frac{\delta f_M}{\bar{f}_M}+\frac{\delta \lambda_{P_c}}{\bar{\lambda}_{P_c}}+\frac{\delta \lambda_{B}}{\bar{\lambda}_{B}}+\frac{\delta G}{\bar{G}}=\frac{\delta f(x_i)}{\bar{f}_M\bar{\lambda}_{P_c}\bar{\lambda}_B\bar{G}},
\end{eqnarray}
where $\bar{f}$ denotes the central value of $f$. For simplicity and also to avoid overestimating uncertainties, we take the approximate relation:
\begin{eqnarray}\label{eq:30}
 \frac{\delta f_M}{\bar{f}_M}\approx\frac{\delta \lambda_{P_c}}{\bar{\lambda}_{P_c}}\approx\frac{\delta \lambda_{B}}{\bar{\lambda}_{B}}\approx\frac{\delta G}{\bar{G}}.
 \end{eqnarray}
 Then, the uncertainty of strong coupling constant can be expressed as,
 \begin{eqnarray}\label{eq:31}
 \delta G=\frac{\delta f(x_i)}{4\bar{f}_M\bar{\lambda}_{P_c}\bar{\lambda}_B}.
 \end{eqnarray}
 Since $f(x_i)$ is a numerical integral of the QCD level input, its partial derivative can not be obtained directly. Therefore, we introduce the following standard deviation formula:
 \begin{eqnarray}\label{eq:32}
 \delta f(x_i)=\sqrt{\sum_{i} \left[f(\bar{x}_i+\delta x_i)-f(\bar{x}_i)\right]^2}.
 \end{eqnarray}
Finally, the uncertainty of strong coupling constant can be obtained as,
\begin{eqnarray}\label{eq:33}
	\delta G=\frac{\sqrt{\sum_{i} \left[f(\bar{x}_i+\delta x_i)-f(\bar{x}_i)\right]^2}}{4\bar{f}_M\bar{\lambda}_{P_c}\bar{\lambda}_B}.
\end{eqnarray}
 
\begin{figure*}
	\centering
	\includegraphics[width=18cm]{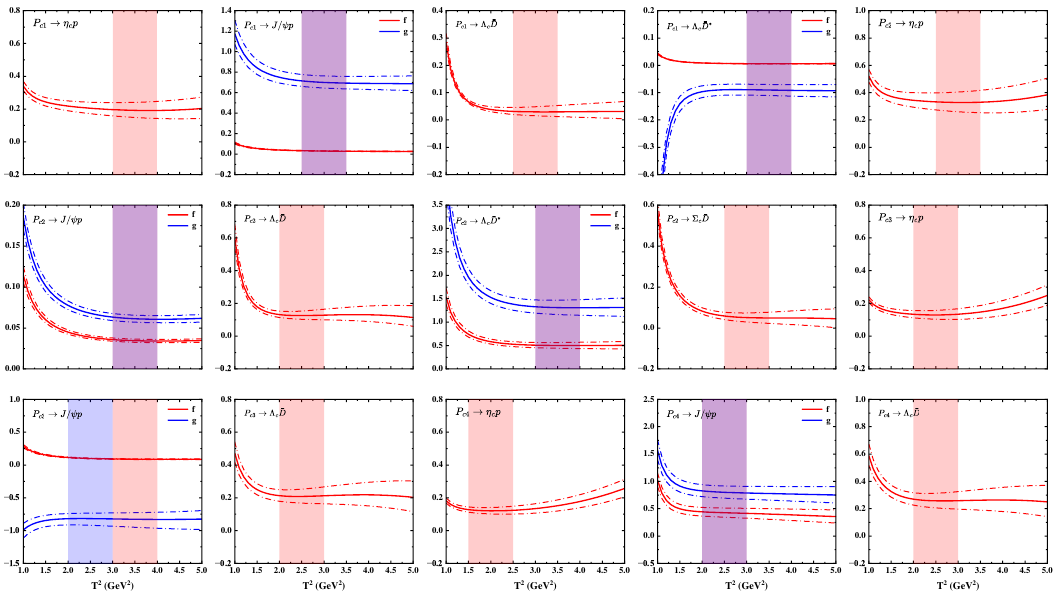}
	\caption{The strong coupling constants of all decay processes with variation of the Borel parameter $T^2$, where the solid and dashed lines represent the central values and the error bands, respectively, and the dark region indicates the Borel platform.}
	\label{BP}
\end{figure*}

The effective subtraction terms $C$ are introduced to absorb the unknown pole-continuum transition contributions and are adjusted to obtain stable Borel platforms. We therefore do not assign independent uncertainties to them. Once stable platforms are obtained, the uncertainties induced by the Borel parameters are small and are included in the above numerical procedure.
The Borel platforms for the all strong coupling constants are collected in Fig.~\ref{BP}. The uncertainties of strong coupling constants are also collected in Table~\ref{tab:2}.

\begin{table*}[htbp]
\begin{ruledtabular}\caption{The effective subtraction terms $C$ and the Borel platforms adopted in the numerical analysis.}\label{tab:2}
\renewcommand\arraystretch{1.3}
\begin{tabular}{l c c c c}
Decay channel & Coupling constant & $C$ & $T^2$ (GeV$^2$)&Values \\
\hline
$P_{c1}\to \eta_c p$ & $G$ & $-9.0\times10^{-6}$ GeV$^{9}$ ($T^2-1.2$ GeV$^2$) & $3.0-4.0$&$0.192^{+0.055}_{-0.034}$ \\
\multirow{2}{*}{$P_{c1}\to J/\psi p$}
& $f$ & $-5.0\times10^{-8}$ GeV$^7$ $T^2$ & $2.5-3.5$&$0.0303^{+0.0013}_{-0.0000}$ \\
& $g$ & $1.1\times10^{-6}$ GeV$^{7}$ $T^2$ & $2.5-3.5$&$0.700^{+0.059}_{-0.039}$ \\
$P_{c1}\to \Lambda_c\bar D$ & $G$ & $-1.8\times10^{-7}$ GeV$^{9}$ ($T^2-2.0$ GeV$^{2}$) & $2.5-3.5$&$0.0297^{+0.0232}_{-0.0094}$ \\
\multirow{2}{*}{$P_{c1}\to \Lambda_c\bar D^{*}$}
& $f$ & $6.5\times10^{-9}$ GeV$^7$ ($T^2-2.4$ GeV$^2$) & $3.0-4.0$&$0.00574^{+0.00051}_{-0.00006}$ \\
& $g$ & $0$ & $3.0-4.0$&$-0.0904^{+0.0212}_{-0.0223}$ \\ \hline
$P_{c2}\to \eta_c p$ & $G$ & $3.2\times10^{-5}$ GeV$^9$ ($T^2-1.2$ GeV$^2$) & $2.5-3.5$&$0.328^{+0.089}_{-0.055}$ \\
\multirow{2}{*}{$P_{c2}\to J/\psi p$}
& $f$ & $6.5\times10^{-8}$ GeV$^7$ $T^2$ & $3.0-4.0$&$0.0345^{+0.0015}_{-0.0006}$ \\
& $g$ & $-7.5\times10^{-8}$ GeV$^7$ $T^2$ & $3.0-4.0$&$0.0613^{+0.0038}_{-0.0022}$ \\
$P_{c2}\to \Lambda_c\bar D$ & $G$ & $1.8\times10^{-6}$ GeV$^9$ $T^2$ & $2.0-3.0$&$0.126^{+0.039}_{-0.012}$ \\
\multirow{2}{*}{$P_{c2}\to \Lambda_c\bar D^{*}$}
& $f$ & $6.2\times10^{-7}$ GeV$^7$ $T^2$ & $3.0-4.0$&$0.502^{+0.069}_{-0.048}$ \\
& $g$ & $1.5\times10^{-6}$ GeV$^7$ $T^2$ & $3.0-4.0$&$1.311^{+0.169}_{-0.118}$ \\
$P_{c2}\to \Sigma_c\bar D$ & $G$ & $2.5\times10^{-5}$ GeV$^9$ $T^2$ & $2.5-3.5$&$0.0508^{+0.0265}_{-0.0096}$ \\ \hline
$P_{c3}\to \eta_c p$ & $G$ & $-1.6\times10^{-5}$ GeV$^9$ ($T^2-1.2$ GeV$^2$) & $2.0-3.0$&$0.130^{+0.038}_{-0.017}$ \\
\multirow{2}{*}{$P_{c3}\to J/\psi p$}
& $f$ & $2.2\times10^{-7}$ GeV$^7$ $T^2$ & $3.0-4.0$&$0.0888^{+0.0042}_{-0.0017}$ \\
& $g$ & $4.0\times10^{-6}$ GeV$^7$ $T^2$ & $2.0-3.0$&$-0.818^{+0.077}_{-0.116}$ \\
$P_{c3}\to \Lambda_c\bar D$ & $G$ & $7.0\times10^{-6}$ GeV$^9$ $T^2$ & $2.0-3.0$&$0.208^{+0.058}_{-0.029}$ \\ \hline
$P_{c4}\to \eta_c p$ & $G$ & $-1.5\times10^{-5}$ GeV$^9$ ($T^2-1.1$ GeV$^2$) & $1.5-2.5$&$0.120^{+0.028}_{-0.008}$ \\
\multirow{2}{*}{$P_{c4}\to J/\psi p$}
& $f$ & $8.0\times10^{-6}$ GeV$^7$ $T^2$ & $2.0-3.0$&$0.427^{+0.080}_{-0.062}$ \\
& $g$ & $-8.5\times10^{-6}$ GeV$^7$ $T^2$ & $2.0-3.0$&$0.800^{+0.109}_{-0.079}$ \\
$P_{c4}\to \Lambda_c\bar D$ & $G$ & $1.2\times10^{-5}$ GeV$^9$ $T^2$ & $2.0-3.0$&$0.258^{+0.067}_{-0.031}$ \\
\end{tabular}
\end{ruledtabular}
\end{table*}

With the estimated strong coupling constants, it is straightforward to calculate the partial decay widths of corresponding decay channels. For the two-body strong decays of the hidden-charm pentaquark states, the standard two-body decay formula can be written as: 
\begin{eqnarray}\label{eq:34}
\notag
\Gamma(P_c\to BM)&=&\frac{1}{2J+1}\sum\frac{p}{8\pi m_{P_c}^2}|\mathcal{M}|^2,\\
\notag
p&=&\frac{\sqrt{[m_{P_c}^2-(m_{B}+m_{M})^2][m_{P_c}^2-(m_{B}-m_{M})^2]}}{2m_{P_c}}.\\
\end{eqnarray}
where $m_{B}$ and $m_{M}$ denote the masses of final baryon and mesons, $J$ is the total angular momentum of the initial $P_c$ states, $\sum$ represents the summation of all the spinor and polarization vectors, and $\mathcal{M}$ is the invariant decay amplitude.

For the decay channels involving pseudoscalar meson, the decay amplitudes are written as
\begin{eqnarray}\label{eq:35}
\mathcal{M}(P_c\to \eta_c p)&=&iG_{P_c\eta_c p}\bar U_p(q,m_p)U_{P_c}(p',m_{P_c}) ,
\nonumber\\
\mathcal{M}(P_c\to \bar D\Lambda_c)&=&iG_{P_c\bar D\Lambda_c}\bar U_{\Lambda_c}(p,m_{\Lambda_c})U_{P_c}(p',m_{P_c}) ,
\nonumber\\
\mathcal{M}(P_c\to \bar D\Sigma_c)&=&iG_{P_c\bar D\Sigma_c}\bar U_{\Sigma_c}(p,m_{\Sigma_c})U_{P_c}(p',m_{P_c}).
\end{eqnarray}
For the decay channels involving vector meson, the decay amplitudes can expressed as
\begin{eqnarray}\label{eq:36}
\notag
\mathcal{M}(P_c\to J/\psi p) &=&\bar U_p(q,m_p)\varepsilon_\alpha^{J/\psi*}(p)\Bigg[f_{P_cJ/\psi p}\gamma^\alpha \\
\notag
&&-\frac{i g_{P_cJ/\psi p}}{m_{P_c}+m_p}\sigma^{\alpha\beta}p_\beta\Bigg]\gamma_5 U_{P_c}(p',m_{P_c}) ,\\
\notag
\mathcal{M}(P_c\to \bar D^*\Lambda_c) &=&\bar U_{\Lambda_c}(p,m_{\Lambda_c})\varepsilon_\alpha^{\bar{D}^{*}*}(q)\Bigg[f_{P_c\bar D^{*}\Lambda_c}\gamma^\alpha \\
\notag
&&-\frac{i g_{P_c\bar D^{*}\Lambda_c}}{m_{P_c}+m_{\Lambda_c}}\sigma^{\alpha\beta}q_\beta\Bigg]\gamma_5 U_{P_c}(p',m_{P_c}) .\\
\end{eqnarray}
The quantities $G$, $f$ and $g$ are the strong coupling constants extracted from the QCD sum rules. According to Eq.~(\ref{eq:34})-(\ref{eq:36}), the partial and total decay widths for these hidden-charm pentaquark are obtained and are listed in Table~\ref{tab:3}. 

A comparison with the LHCb measurements reveals that the mass and total width of our $P_{c1}$ state are $4310\pm110$ and $13.758^{+2.502}_{-1.568}$ MeV, which are in agreement with the observed $P_c(4312)$, whose mass and width are $4311.9\pm0.7^{+6.8}_{-0.6}$ and $9.8\pm2.7^{+3.7}_{-4.5}$ MeV, respectively~\cite{LHCb:2019kea}. This consistency supports the interpretation of $P_c(4312)$ as a compact pentaquark state described by the interpolating current $J_1$ in Eq.~(\ref{eq:1}). For the $P_{c2}$ state, its mass lies close to both $P_c(4440)$ and $P_c(4457)$, while its width is $6.995^{+2.711}_{-1.321}$ MeV, which is particularly close to that of $P_c(4457)$ ($6.4\pm2.0^{+5.7}_{-1.9}$ MeV)~\cite{LHCb:2019kea}. Therefore, we assign $P_c(4457)$ as the more plausible counterpart of $P_{c2}$, and identify it as a compact pentaquark state generated by the current $J_2$. Furthermore, we predict two additional compact pentaquark states $P_{c3}$ and $P_{c4}$ with masses of $4200\pm110$ and $4250\pm110$ MeV and total widths of $22.429^{+5.694}_{-3.323}$ and $3.420^{+1.627}_{-0.626}$ MeV, respectively. Their dominant decay modes are predicted to be $J/\psi p$ and $\Lambda_c\bar{D}$, respectively. We suggest that future experiments search for these two new states in the invariant mass distributions of the corresponding final states.
\begin{table*}[htbp]
\begin{ruledtabular}\caption{The partial and total decay widths of the hidden-charm pentaquark states. All quantities are in units of MeV.}\label{tab:3}
\renewcommand{\arraystretch}{1.5}
\begin{tabular}{c c c c c c c c c}
States
& Mass & $\eta_c p$ & $J/\psi p$ & $\Lambda_c\bar D$ & $\Lambda_c\bar D^{*}$ & $\Sigma_c\bar D$ & Total &Assignment\\
\hline
$P_{c1}$
& $4310\pm110$ & $1.133^{+0.737}_{-0.362}$ & $12.530^{+2.388}_{-1.525}$ & $0.045^{+0.098}_{-0.024}$ & $0.049^{+0.024}_{-0.017}$ & $-$ & $13.758^{+2.502}_{-1.568}$&$P_c(4312)$ \\
$P_{c2}$
& $4450\pm110$ & $4.111^{+2.528}_{-1.255}$ & $0.002^{+0.001}_{-0.000}$ & $1.076^{+0.760}_{-0.190}$ & $1.686^{+0.594}_{-0.364}$ & $0.120^{+0.158}_{-0.041}$ & $6.995^{+2.711}_{-1.321}$&$P_c(4457)$ \\
$P_{c3}$
& $4200\pm110$ & $0.438^{+0.294}_{-0.109}$ & $20.709^{+5.627}_{-3.304}$ & $1.282^{+0.818}_{-0.337}$ & $-$ & $-$ & $22.429^{+5.694}_{-3.323}$&$?$ \\
$P_{c4}$
& $4250\pm110$ & $0.410^{+0.214}_{-0.053}$ & $0.293^{+0.106}_{-0.074}$
& $2.718^{+1.610}_{-0.619}$ & $-$ & $-$ & $3.420^{+1.627}_{-0.626}$&$?$ \\
\end{tabular}
\end{ruledtabular}
\end{table*}

The numerical results in Table~\ref{tab:3} show that the decay behaviors depend strongly on the interpolating currents. For the current $J_1$, the $J/\psi p$ channel gives the dominant contribution to the total width, while the $\eta_c p$, $\Lambda_c\bar D$ and $\Lambda_c\bar D^{*}$ channels are suppressed. For the current $J_2$, several open-charm channels contribute to the total width, and the resulting state remains relatively narrow. For the currents $J_3$ and $J_4$, the $J/\psi p$ and $\Lambda_c\bar D$ channels provide the main contributions. These different decay behaviors may be helpful for distinguishing these states in future experiments. 

With the partial decay widths obtained above, we can analyze the relative importance of different decay channels. For this purpose, we define the branching fraction as
\begin{eqnarray}
{\cal B}(P_c\to BM)=\frac{\Gamma(P_c\to BM)}{\Gamma_{\rm total}}.
\end{eqnarray}
\begin{table}[htbp]
\begin{ruledtabular}\caption{The central values of branching fraction of the hidden-charm pentaquark states.}\label{tab:4}
\begin{tabular}{c c c c c c}
States
& $\eta_c p$ & $J/\psi p$ & $\Lambda_c\bar D$ & $\Lambda_c\bar D^{*}$ & $\Sigma_c\bar D$ \\
\hline
$P_{c1}$
& $0.0824$ & $0.911$ & $3.271\times10^{-3}$ & $3.562\times10^{-3}$ & $-$ \\
$P_{c2}$
& $0.588$ & $0.286\times10^{-3}$ & $0.154$ & $0.241$ & $0.0172$ \\
$P_{c3}$
& $0.0195$ & $0.923$ & $0.0572$ & $-$ & $-$ \\
$P_{c4}$
& $0.120$ & $0.0857$ & $0.795$ & $-$ & $-$ \\
\end{tabular}
\end{ruledtabular}
\end{table}
The total width $\Gamma_{\rm total}$ is obtained by summing over all partial decay widths considered in this work, and the corresponding branching fractions are listed in Table~\ref{tab:4}. Theoretically, there remains considerable controversy regarding the partial decay widths of these $P_c$ states. For $P_c(4312)$, most studies interpret it as a $\Sigma_c\bar{D}$ molecular state with $I(J^P)$ of $\frac{1}{2}(\frac{1}{2}^-)$. In recent years, several groups have investigated its partial decay widths within the molecular framework using different approaches. In Ref.~\cite{Xu:2020flp}, the two-body strong decays of $P_c(4312)$ were studied using the three-point QCD sum rules, revealing that the hidden-charm channels $\eta_cp$ and $J/\psi p$ dominate, with a branching ratio $\frac{\mathcal{B}(P_c(4312)\to\eta_cp)}{\mathcal{B}(P_c(4312)\to J/\psi p)}\approx3.3$. However, a similar study based on effective Lagrangians and triangle diagrams in Ref.~\cite{Deng:2026gqe} suggests that the open-charm decay modes, particularly $\Lambda_c\bar{D}^*$, are dominant. Our previous QCD sum rule analysis~\cite{Wang:2023ews} also found that hidden-charm channels dominate, but yielded a much smaller ratio of $\frac{\mathcal{B}(P_c(4312)\to\eta_cp)}{\mathcal{B}(P_c(4312)\to J/\psi p)}\approx0.01$. In the present compact-pentaquark framework, as shown in Table~\ref{tab:4}, the open-charm decays of 
$P_c(4312)$ are significantly suppressed, while the dominant modes are hidden-charm channels, with a ratio $\frac{\mathcal{B}(P_c(4312)\to\eta_cp)}{\mathcal{B}(P_c(4312)\to J/\psi p)}\approx0.09$. For $P_c(4457)$, most theoretical interpretations favour a $\Sigma_c^{(*)}\bar{D}^*$ molecular assignment with $I(J^P)=\frac{1}{2}(\frac{3}{2}^-)$ or $\frac{1}{2}(\frac{5}{2}^-)$~\cite{Wu:2019rog,Xiao:2019mvs,Wang:2023ews}. Nevertheless, Ref.~\cite{Deng:2026gqe} supports a low-spin $J=\frac{1}{2}$ assignment, which is consistent with our conclusion. It is worth noting that in their results the hidden-charm channels are strongly suppressed, and the decays proceed predominantly through open-charm modes, whereas in our results the hidden-charm channel $\eta_c p$ dominates. This distinct pattern may serve as a useful discriminant between compact and molecular configurations. In summary, theoretical predictions for the observed $P_c$ states exhibit substantial discrepancies, underscoring the need for further experimental scrutiny. 
\section{Conclusion}\label{sec4}
In the present work, we have studied the two-body strong decays of the hidden-charm pentaquark states coupled to four local interpolating currents with the quantum numbers $I(J^P)=\frac{1}{2}(\frac{1}{2}^{-})$ in the framework of the three-point QCD sum rules. The strong coupling constants are extracted from the selected Lorentz structures, and the corresponding partial decay widths are calculated with the obtained coupling constants. The numerical results indicate that the state coupled to $J_1$ may be assigned to the $P_c(4312)$, while the state coupled to $J_2$ may be related to the $P_c(4457)$, as both the masses and widths are compatible with the experimental data within uncertainties.

It is worth comparing the present compact pentaquark assignments with the molecular interpretations of the $P_c(4312)$ and $P_c(4457)$ proposed in previous studies~\cite{Wang:2023ews,Wang:2026bls}. In the molecular picture, the proximity of the observed masses to the corresponding charmed-baryon and anti-charmed-meson thresholds plays an essential role. In the present diquark-diquark-antiquark picture, however, the masses and decay widths can also be reproduced within uncertainties by using local currents with definite isospin. Therefore, the decay patterns, especially the relative strengths of the $J/\psi p$, $\eta_c p$ and open charm channels, are expected to provide useful information for distinguishing compact pentaquark configurations from hadronic molecular configurations.

The states coupled to the currents $J_3$ and $J_4$ are predicted to be possible hidden-charm pentaquark candidates with lower masses. Their dominant decay channels are $J/\psi p$ and $\Lambda_c\bar D$, which may serve as useful experimental signatures. Searching for such structures in the corresponding invariant mass distributions will be helpful for testing the compact pentaquark assignments and for understanding the spectrum and decay mechanism of the hidden-charm pentaquark states.

As a natural extension, the same three-point QCD sum rule analysis can be applied to the hidden-charm pentaquark currents with $I(J^P)=\frac{1}{2}(\frac{3}{2}^{-})$ and $\frac{1}{2}(\frac{5}{2}^{-})$ constructed in Ref.~\cite{Wang:2025qtm}. The calculation of their strong coupling constants and two-body decay widths will make it possible to compare the $\frac{1}{2}^{-}$, $\frac{3}{2}^{-}$ and $\frac{5}{2}^{-}$ compact pentaquark assignments in a unified framework, and may be helpful for clarifying the nature of the $P_c(4337)$, $P_c(4380)$, $P_c(4440)$ and $P_c(4457)$ states.

\section*{Acknowledgements}
This work is supported by National Natural Science Foundation,
Grant No. 12575083, and Natural Science Foundation of HeBei Province under the Grant No. A2024502002.

\bibliographystyle{apsrev4-1}
\bibliography{refs}

\end{document}